\documentclass[a4paper,11pt]{article}
\pdfoutput=1 % if your are submitting a pdflatex (i.e. if you have
\usepackage{jinstpub} % for details on the use of the package, please
\usepackage[acronym]{glossaries-extra}
\setabbreviationstyle[acronym]{long-short}
\glssetcategoryattribute{acronym}{nohyper}{true}
\newacronym{da}{DA}{Deconvolution Algorithm}
\newacronym{snr}{SNR}{Signal-to-Noise Ratio}
\newacronym{psd}{PSD}{Power Spectral Density}
\newacronym{spsd}{SPSD}{Signal Power Spectral Density}
\newacronym{npsd}{NPSD}{Noise Power Spectral Density}
\newacronym{ifd}{IFD}{Inverse Filtering Deconvolution}
\newacronym{lti}{LTI}{Linear and Time Invariant}
\newacronym{dsp}{DSP}{Digital Signal Processor}
\newacronym{fir}{FIR}{Finite Impulse Response}
\newacronym{fpga}{FPGA}{Field Programmable Gate Array}
\newacronym{dtft}{DTFT}{Discrete-time Fourier Transform}
\newacronym{idtft}{IDTFT}{Inverse Discrete-time Fourier Transform}
\newacronym{dft}{DFT}{Discrete Fourier Transform}
\newacronym{fft}{FFT}{Fast Fourier Transform}
\newacronym{ifft}{IFFT}{Inverse Fast Fourier Transform}
\newacronym{juno}{JUNO}{Jiangmen Underground Neutrino Observatory}
\newacronym{adc}{ADC}{Analog-to-Digital Converter}
\newacronym{fadc}{Flash-ADC}{Flash Analog-to-Digital Converter}
\newacronym{lpmt}{LPMT}{Large Photo Multiplier Tube}
\newacronym{spmt}{SPMT}{Small Photo Multiplier Tube}
\newacronym{pmt}{PMT}{Photo Multiplier Tube}
\newacronym{pe}{PE}{photoelectron}
\newacronym{spe}{SPE}{single photoelectron}
\newacronym{gcu}{GCU}{Global Control Unit}
\newacronym{daq}{DAQ}{Data Acquisition System}
\newacronym{wf}{WF}{waveform}
\newacronym{tq}{TQ}{time-charge pair}
\newacronym{coti}{COTI}{Continuous Over Threshold Integration}
\newacronym{catiroc}{CATIROC}{Charge And Time Integrated Read Out Chip}
\newacronym{bibo}{BIBO}{Bounded-Input, Bounded-Output}
\newacronym{lmse}{LMSE}{Least Mean Squared Error}
\newacronym{cd}{CD}{Central Detector}
\newacronym{serdes}{SERDES}{Serializer-Deserializer}
\newacronym{hk}{Hyper-K}{Hyper-Kamiokande}
\newacronym{rtwd}{RTWD}{Real-Time Wiener Deconvolution}
\newacronym{ctu}{CTU}{Central Trigger Unit}
\newacronym{mcp}{MCP}{Micro Channel Plate}
\newacronym{rms}{RMS}{Root Mean-Squared}
\newacronym{rtl}{RTL}{Register Transfer level}
\newacronym{lut}{LUT}{Look-Up Table}
\newacronym{lutram}{LUTRAM}{Look-Up Table RAM}
\newacronym{bram}{BRAM}{Block RAM}
\newacronym{ff}{FF}{Flip-Flop}
\usepackage{orcidlink}
\usepackage{tikz}
\usetikzlibrary{shapes, arrows.meta, positioning, calc, fit}
\usepackage{tabularray}

\title{\boldmath Real-Time Wiener Deconvolution for feature reconstruction in JUNO}

\author[a,1]{L.~Lastrucci${}^{\orcidlink{0009-0002-9838-4593}}$\note{Corresponding Author}}
\emailAdd{lorenzolastrucci1@gmail.com}

\author[a,1]{M.~Grassi${}^{\orcidlink{0000-0003-2422-6736}}$}
\emailAdd{marco.grassi@unipd.it}

\author[a]{A.~Triossi${}^{\orcidlink{0000-0001-5140-9154}}$}

\author[b,c]{J.~Hu}

\author[b,c]{X.~Jiang}

\author[a]{R.~Brugnera${}^{\orcidlink{0000-0002-2115-3992}}$}

\author[a]{A.~Garfagnini${}^{\orcidlink{0000-0003-0658-1830}}$}

\author[a]{V.~Cerrone${}^{\orcidlink{0000-0002-1434-8804}}$}

\author[a]{L.~V.~D'Auria${}^{\orcidlink{0009-0004-8774-5140}}$}

\author[a]{A.~Gavrikov${}^{\orcidlink{0000-0002-6741-5409}}$}

\author[a]{R.~M.~Guizzetti${}^{\orcidlink{0009-0008-5007-4506}}$}

\author[a]{A.~Serafini${}^{\orcidlink{0000-0001-9191-661X}}$}

\author[d]{G.~Andronico${}^{\orcidlink{0000-0002-0079-5021}}$}

\author[e]{V.~Antonelli${}^{\orcidlink{0000-0001-9680-0149}}$}

\author[f]{A.~Barresi${}^{\orcidlink{0000-0002-6859-0903}}$}

\author[e]{D.~Basilico${}^{\orcidlink{0000-0001-5662-9236}}$}

\author[e]{M.~Beretta${}^{\orcidlink{0009-0009-7114-5200}}$}

\author[a]{A.~Bergnoli}

\author[f]{M.~Borghesi${}^{\orcidlink{0000-0001-5854-8894}}$}

\author[e]{A.~Brigatti}

\author[d]{R.~Bruno}

\author[g]{A.~Budano${}^{\orcidlink{0000-0002-0856-1131}}$}

\author[e]{B.~Caccianiga${}^{\orcidlink{0000-0002-8026-7754}}$}

\author[h]{A.~Cammi${}^{\orcidlink{0000-0003-1508-5935}}$}

\author[d]{R.~Caruso${}^{\orcidlink{0000-0003-1622-8731}}$}

\author[f]{D.~Chiesa${}^{\orcidlink{0000-0003-1978-1727}}$}

\author[i]{C.~Clementi${}^{\orcidlink{0000-0003-0470-2870}}$}

\author[f]{C.~Coletta${}^{\orcidlink{0009-0004-6004-8038}}$}

\author[a]{S.~Dusini${}^{\orcidlink{0000-0002-1128-0664}}$}

\author[g]{A.~Fabbri${}^{\orcidlink{0000-0002-6447-9968}}$}

\author[j]{G.~Felici${}^{\orcidlink{0000-0001-8783-6115}}$}

\author[f]{G.~Ferrante${}^{\orcidlink{0009-0002-2741-0811}}$}

\author[e]{M.~G.~Giammarchi${}^{\orcidlink{0000-0002-6509-6467}}$}

\author[d]{N.~Giudice}

\author[d]{N.~Guardone}

\author[e]{F.~Houria${}^{\orcidlink{0009-0009-7458-345X}}$}

\author[g]{A.~Islam}

\author[e]{C.~Landini${}^{\orcidlink{0000-0002-4717-488X}}$}

\author[a]{I.~Lippi${}^{\orcidlink{0000-0002-8181-3905}}$}

\author[h]{L.~Loi${}^{\orcidlink{0009-0003-7904-3731}}$}

\author[e]{P.~Lombardi${}^{\orcidlink{0000-0003-0853-3154}}$}

\author[k]{F.~Mantovani${}^{\orcidlink{0000-0003-1200-0174}}$}

\author[g]{S.~M.~Mari${}^{\orcidlink{0000-0002-5973-5103}}$}

\author[j]{A.~Martini${}^{\orcidlink{0000-0001-8909-8048}}$}

\author[e]{L.~Miramonti${}^{\orcidlink{0000-0002-2808-5363}}$}

\author[k]{M.~Montuschi${}^{\orcidlink{0000-0002-2930-5030}}$}

\author[f]{M.~Nastasi${}^{\orcidlink{0000-0001-7967-999X}}$}

\author[g]{D.~Orestano${}^{\orcidlink{0000-0001-5103-5527}}$}

\author[i]{F.~Ortica${}^{\orcidlink{0000-0001-8276-452X}}$}

\author[j]{A.~Paoloni${}^{\orcidlink{0000-0002-4141-7799}}$}

\author[e]{L.~Pelicci${}^{\orcidlink{0000-0001-8283-3388}}$}

\author[e]{E.~Percalli${}^{\orcidlink{0009-0007-7241-325X}}$}

\author[g]{F.~Petrucci${}^{\orcidlink{0000-0002-5278-2206}}$}

\author[f]{E.~Previtali${}^{\orcidlink{0000-0003-0028-718X}}$}

\author[e]{G.~Ranucci${}^{\orcidlink{0000-0002-3591-8191}}$}

\author[e]{A.~C.~Re${}^{\orcidlink{0000-0002-2340-7802}}$}

\author[k]{B.~Ricci${}^{\orcidlink{0000-0001-6161-4098}}$}

\author[i]{A.~Romani${}^{\orcidlink{0000-0002-7338-0097}}$}

\author[a]{C.~Sirignano${}^{\orcidlink{0000-0002-0995-7146}}$}

\author[f]{M.~Sisti${}^{\orcidlink{0000-0003-2517-1909}}$}

\author[a]{L.~Stanco${}^{\orcidlink{0000-0002-9706-5104}}$}

\author[g]{E.~Stanescu~Farilla${}^{\orcidlink{0009-0007-2472-5515}}$}

\author[k]{V.~Strati${}^{\orcidlink{0000-0001-7271-5353}}$}

\author[e]{M.~D.~C.~Torri${}^{\orcidlink{0000-0002-8022-3495}}$}

\author[d]{C.~Tuvè${}^{\orcidlink{0000-0003-0739-3153}}$}

\author[g]{C.~Venettacci${}^{\orcidlink{0009-0009-3009-9797}}$}

\author[d]{G.~Verde${}^{\orcidlink{0000-0002-8622-8297}}$}

\author[j]{L.~Votano${}^{\orcidlink{0000-0002-1772-1328}}$}

\author[l]{G.~Dong}

\author[l]{J.~Dong}

\author[b,c]{L.~Fan}

\author[b]{S.~Hou}

\author[b,c]{Z.~Ning}

\author[b]{Y.~Sun}

\author[b]{Y.~Wang}

\author[b,c]{Z.~Wang}

\author[b,c]{X.~Yan}

\affiliation[a]{INFN, Sezione di Padova e Università di Padova, Dipartimento di Fisica e Astronomia, Padua, Italy}
\affiliation[b]{Institute of High Energy Physics, Chinese Academy of science}
\affiliation[c]{University of Chinese Academy of science}
\affiliation[d]{INFN, Sezione di Catania e  Università di Catania, Dipartimento di Fisica e Astronomia, Catania, Italy}
\affiliation[e]{INFN, Sezione di Milano e Università degli Studi di Milano, Dipartimento di Fisica, Milan, Italy}
\affiliation[f]{INFN, Sezione di Milano Bicocca e Dipartimento di Fisica Università di Milano Bicocca, Milan, Italy}
\affiliation[g]{INFN, Sezione di Roma Tre e Università degli Studi Roma Tre, Dipartimento di Matematica e Fisica, Rome, Italy}
\affiliation[h]{INFN, Sezione di Milano Bicocca e Dipartimento di Energetica, Politecnico di Milano, Milan, Italy}
\affiliation[i]{INFN, Sezione di Perugia e Università degli Studi di Perugia, Dipartimento di Chimica, Biologia e Biotecnologie, Perugia, Italy}
\affiliation[j]{Laboratori Nazionali dell'INFN di Frascati, Frascati, Italy}
\affiliation[k]{INFN, Sezione di Ferrara, Ferrara, Italy e Università degli Studi di Ferrara, Dipartimento di Fisica e Scienze della Terra, Ferrara, Italy}
\affiliation[l]{Tsinghua University}

\abstract{In particle physics, experiments generate substantial amounts of data that can be difficult to process without preliminary scaling. To avoid losing potentially crucial data, experimental collaborations are studying novel techniques for real-time data processing to extract features for further physics analysis. A common approach, especially in neutrino physics, is to use FPGAs for data acquisition and pre-processing. 
This paper presents an advanced Real-Time Wiener deconvolution algorithm designed to leverage the processing capabilities of the FPGA integrated into the readout boards of the Jiangmen Underground Neutrino Observatory (JUNO). The goal is to enable real-time reconstruction of the signal generated by photomultiplier tubes (PMTs) when neutrino interactions are detected. 
By exploiting online reconstruction of the signal generated by PMTs, we expect to improve the detection of low-energy depositions, such as those produced by transient astrophysical phenomena. These depositions are usually not saved because of the significant background  that affects the low end of the energy spectrum, which would result in a large trigger rate, hence a large amount of data required for storage. 
This paper presents the features of the algorithm, including its ability to manage high-throughput data streams with minimal latency, adaptability, and resilience in discerning the characteristics of input data. Performance is evaluated on a JUNO electronic board. This study further demonstrates the potential of FPGA-based solutions for neutrino physics.}


\begin{document}
\maketitle
\flushbottom
\section{Introduction}
\label{sec:intro}
In experimental particle physics, there is a growing trend toward collecting and analyzing large volumes of data. This is particularly relevant at the intensity frontier, where large detector volumes are often required to compensate for low cross-sections and low fluxes. Neutrino experiments based on \gls{pmt} technology are a prime example, where large detection volumes correspond to a high count of readout channels. To handle these signals while preserving the features of the \gls{pmt} output, recent and future experiments~\cite{An_2016,LAVITOLA2023168461,IceCube:2016zyt,KM3NeT:2022pnv} implement analog-to-digital conversion  near the \gls{pmt}.

The transmission of digital data to the \gls{daq} farm requires low-level processing, usually achieved using a \gls{fpga}. \glspl{fpga} offer significant computational capabilities, re-programmability, and operational consistency, all while maintaining relatively low costs.

Beyond handling I/O, \gls{fpga} resources can perform signal pre-processing and filtering. This enables the extraction of relevant features from the \gls{pmt} output signal (hereafter referred to as the \textit{waveform}), avoiding the need to transmit the entire sampled waveform—a task that can be extremely demanding given high sampling frequencies and large channel counts. This approach is crucial when an experiment searches for multiple signatures that differ in rate and energy-resolution requirements. Events requiring high energy resolution benefit from high-frequency digitization to properly disentangle all photons hitting the \gls{pmt}. However, the acquisition rate for such events is limited by the bandwidth between the readout electronics and the \gls{daq}. To retain high-rate signatures, such as low energy events highly contaminated by natural radioactivity background, online pre-processing must occur on the readout electronics board, ensuring that only the most relevant waveform features are transmitted.

In this article, we present a novel approach to waveform feature reconstruction called \gls{rtwd}, a real-time reconstruction technique for \gls{pmt} waveform processing based on the Deconvolution algorithm. It aims to disentangle \gls{spe} contributions ---which we call \textit{hits}--- within a given waveform to improve the reconstruction of hit times and total charge.

This work was initiated in the context of the \gls{juno} experiment; therefore, we describe the \gls{juno} detector and the method's application to it. However, the algorithm's performance and the results of the study are pertinent to any \gls{pmt}-based detector featuring an \gls{fpga} on its readout electronics.

The paper is organized as follows: \autoref{sec:juno} introduces the \gls{juno} experiment, \autoref{sec:setup} describes the experimental setup used to develop and test the algorithm, \autoref{sec:algorithm} introduces the \glsxtrlong{rtwd} algorithm, while \autoref{sec:rtwd} details the \gls{juno} use case. In \autoref{sec:result} we presente the results of the tests conducted. Finally, we draw our conclusions in \autoref{sec:conclusion}.

\glsreset{rtwd}
\section{Jiangmen Underground Neutrino Observatory}
\label{sec:juno}
The \gls{juno} \cite{An_2016} is a 20 kton multipurpose underground liquid scintillator detector designed to study neutrino interactions by detecting scintillation and Cherenkov light. It employs a \textit{dual} photodetection system consisting of 17,612 20-inch \glspl{lpmt} and 25,600 3-inch \glspl{spmt}. The \glspl{lpmt} operate within a wide dynamic range, spanning from a single photoelectron to several hundred \glspl{pe} within a $1~\mu s$ readout window. Each group of three \glspl{lpmt} connects to a readout board known as a \gls{gcu}, which comprises six \glspl{fadc} units and an AMD Kintex 7 \gls{fpga}. As the \glspl{spmt} use a different readout strategy, they are excluded from this study; consequently, the term \gls{pmt} will hereafter be used exclusively to refer to the \gls{lpmt}.

Each \gls{pmt} connects to two amplifiers with distinct gains, producing two output signals per \gls{pmt}. Each signal is sampled at 1 GS/s and digitized with a resolution of 14 bits using a dedicated \gls{fadc}. Every waveform sample is stored as two 8-bit words. The \gls{fadc} digital output is parallelized into eight data streams at 125 MHz via a \gls{serdes} to align with the \gls{fpga} clock. Consequently, for a given \gls{fadc}, the \gls{fpga} processes eight waveform samples per clock cycle---each sample being 16 bits wide---resulting in a total bus width of 128 bits. The primary tasks of the \gls{fpga} are: (i) to receive digitized data from the \gls{fadc}, to pack them, and to transmit them to the \gls{daq}; and (ii) to generate trigger primitives when \gls{pmt} waveforms exceed a defined threshold. These primitives are subsequently evaluated by the Back-End electronics to issue a global trigger, which initiates the readout of all active channels. A more comprehensive description of \gls{gcu} operations and JUNO electronics is provided in ref.~\cite{fmarini}.

Since \gls{juno} \gls{fpga} is responsible for waveform acquisition, trigger generation, and basic data processing, a large fraction of its resources have already been allocated. The goal of this study is to improve trigger efficiency, hit identification, and total charge reconstruction using the limited \gls{fpga} resources that remain available. The baseline operation of the electronics yields two distinct data streams: (i) \textit{\glspl{wf}}, representing a stream of fully digitized \gls{pmt} signals segmented into readout windows—each window comprising  1008 samples—and recorded only upon event validation by the \gls{ctu}; and (ii) \textit{\glspl{tq}}, a trigger-less stream of Time and Charge pairs for each hit, which are continuously reconstructed and saved. 
%The \glspl{wf} represent the full readout window acquired when a \gls{pe} hits a \gls{pmt}. 
The readout system stores the full WF only when the \gls{ctu} issues a \textit{global trigger}, which means that the number of trigger primitives, namely PMTs firing, is above a predefined threshold. 
%validates an event to allow offline reprocessing using precise reconstruction methods. 
Storing all \glspl{wf}—each worth 2 kB—for all events would be infeasible, because the trigger rate induced by cosmic muons is roughly 4 Hz, and storing roughly 17,000 \glspl{wf} per event yields 33.2 MB/event, i.e., approximately 11 TB per day.
% Storing as the data volume would reach approximately 33.2 MB/event (2 KB per waveform times roughly 17,000 channels).
% %; each waveform constitutes a 2 KB packet, and with roughly 17,000 waveforms per event, the data volume reaches approximately 33.2 MB/event. 
% Given the irreducible cosmic muon background trigger rate of $\sim$4 Hz, this would result in roughly 11 TB of data per day.
In contrast, \glspl{tq} serve as a compression of \glspl{wf}. Whenever a \gls{pe} is detected, its charge and timestamp are calculated (currently through the \gls{coti} algorithm, described later in this section) and sent directly to the \gls{daq}. This allows continuous recording of single \gls{pmt} activity—even in the absence of a globally validated event—without exceeding bandwidth limits.
The \gls{coti} method defines a hit whenever the amplitude of the waveform exceeds a threshold and computes its charge as the integral of all samples exceeding the said threshold. Each hit yields a \gls{tq} pair, where the timestamp is defined as the time the threshold was exceeded.
%is based on the integration of digital signals. Its performance relies on the over-threshold detection of hits, where charge is computed 
%However, whenever the waveform exceeds this threshold, the \gls{coti} algorithm generates a \gls{tq} even if the peak was not caused by a \gls{pe} hit. 
The algorithm is able to disentangle two photoelectrons close in time only if the waveform goes back below the threshold between the first and second hit. Consequently, if a group of hits is too close in time, i.e., if the time distance between two hits is shorter than the width of the pulse generated by a single hit (roughly 40 ns in this study), the pileup can cause the \gls{coti} algorithm to merge them into a single TQ pair, thereby degrading the reconstruction performance. This is \gls{coti}'s main limitation, which directly affects the energy evaluation of any interaction taking place in the detector. The \gls{tq} reconstruction algorithm presented in this paper is intended to address primarily the limitation mentioned above.

\glsreset{rtwd}
\section{Experimental setup}
\label{sec:setup}
\begin{figure}[htbp]
    \centering
    \includegraphics[width=0.7\linewidth]{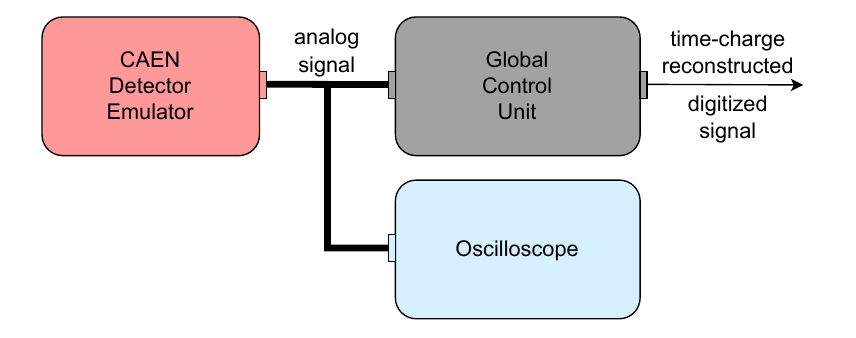}
    \caption{Experimental setup. The experimental setup is composed by a CAEN detector emulator, able to simulate arbitrary \gls{pmt} output signal, a \gls{gcu} connected to the emulator via a coaxial cable and an oscilloscope connected in parallel to monitor the signal shape.}
    \label{fig:exp_setup}
\end{figure}

To validate the performance of the \gls{rtwd} algorithm, we designed an experimental setup (\autoref{fig:exp_setup}) with well-controlled conditions, enabling effective debugging while remaining representative of a \gls{pmt}-based detector like \gls{juno}. 

\begin{figure}[htbp]
    \centering
    \includegraphics[width=0.8\linewidth]{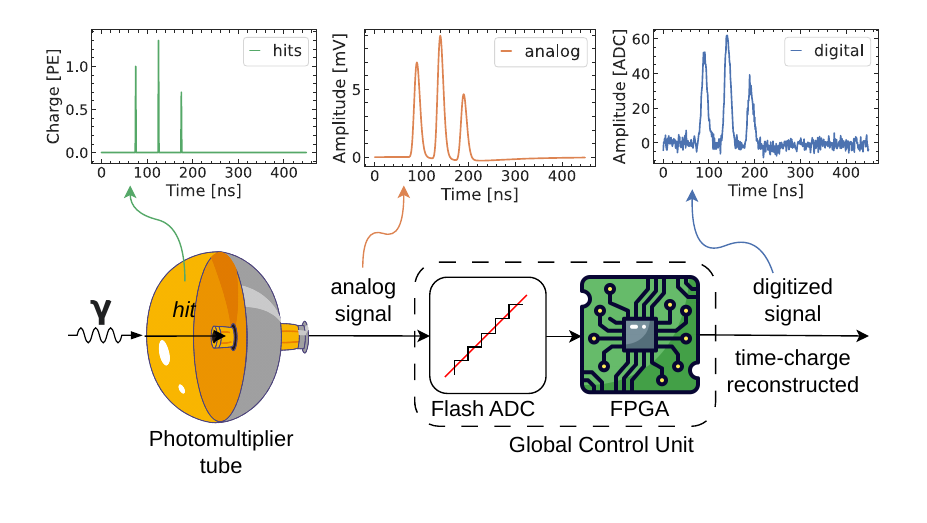}
    \caption{Data acquisition chain. The data acquisition chain can be schematized as follows: a \gls{pmt} detects light and accumulates charge (green); the analog output signal (orange) is digitized by a Flash \gls{adc} and processed by the \gls{fpga}; the digitized signal (blue) and the reconstructed features (time and charge) of the hits are sent to the \gls{daq}. The vertical range of the last two panels is normalized to ease the comparison between the analog and the digital signals.}
    \label{fig:pmtscheme}
\end{figure}

The primary objective was to assess the ability of the algorithm to correctly reconstruct the time and charge of \textit{hits}, charge spikes (green in \autoref{fig:pmtscheme}) in the \gls{pmt} cathode identifiable as over-threshold excursions of the \gls{pmt} output signal. Given the difficulty of precisely controlling the number and intensity of hits in a physical \gls{pmt}, we emulated the detector behavior using the CAEN Fast Digital Detector Emulator DT5810 \cite{caen}. This device generates an analog output that can be configured in amplitude and shape and can be used to simulate arbitrary \gls{pmt} signals to study the behavior and performance of the \gls{rtwd} algorithm. Data acquisition was performed using the final version of the \gls{gcu} \cite{fmarini}, the standard acquisition board used by the \gls{juno} experiment. The digitized \gls{pmt} signal (blue in \autoref{fig:pmtscheme}) and the features extracted by the algorithm are then transmitted to a desktop PC equipped with software inspired by the \gls{juno} \gls{daq}, designed to handle acquisition and data storage. In parallel to this acquisition chain, an oscilloscope has been set to monitor the shape of the detector emulator output. Finally, the data are processed using custom Python libraries to analyze the results. 

\glsreset{rtwd}
\section{Real-Time Wiener Deconvolution}
\label{sec:algorithm}
When the output signal of a \gls{pmt} exceeds a threshold, we define the resulting charge as a \textit{hit}. The objective of this study is to reconstruct this hit charge in real-time on an \gls{fpga}. An effective approach is to employ Wiener Deconvolution~\cite{wiener}, a method based on the Wiener filter (or optimum filter) designed to maximize the \gls{snr} of a signal when both signal and noise characteristics are known.

To minimize the \gls{fpga} resources required for this task, we utilize a \gls{fir} implementation of Wiener Deconvolution, hereafter referred to as \gls{rtwd}. The adoption of \gls{fir} filters is driven by the need for a low-latency algorithm with minimal resource consumption. Due to limitations on local storage and Flash \gls{adc} sampling frequencies ranging from $10^8$ to $10^9$~S/s, the \gls{fpga} must process data with a latency of the order of $100$~ns. Furthermore, neutrino experiments generally employ a vast array of \glspl{pmt} to achieve the necessary photo-coverage, which requires a correspondingly large number of \glspl{fpga}. This scale forces a trade-off between cost and computational power; consequently, the majority of \gls{fpga} resources are allocated to data acquisition, leaving a limited budget for advanced signal processing.

The implementation of \gls{rtwd} comprises four steps: (i) Estimation of the \gls{spe} template; (ii) Characterization of the noise; (iii) Computation of the filter frequency response; and (iv) Derivation of the \gls{fir} filters.

\subsection{Single photoelectron template estimation}
\label{sec:spsd}
The first and most critical step is to obtain a robust characterization of the acquisition system. In this context, \textit{system} refers to all elements that influence the signal of interest, including the processing apparatus and the acquisition instruments. For complex detectors such as \gls{juno}
% or \gls{hk}
, this presents a challenge, as each \gls{pmt} exhibits a unique response. Moreover, different \gls{pmt} types are often combined; for instance, the \gls{juno} Central Detector utilizes approximately 75\% \gls{mcp} \glspl{pmt} from NNVT and 25\% Dynode \glspl{pmt} from Hamamatsu~\cite{Abusleme2022}.

Since the scope of this work is to develop the reconstruction algorithm, we assume to a first approximation that all \glspl{lpmt} exhibit similar behavior, characterizing them with a single \textit{template}. However, the algorithm is natively capable of handling distinct templates for individual channels. Future implementations should include an independent characterization of each \gls{pmt} to further enhance algorithm performance.

\begin{figure}[htbp]
    \centering
    \includegraphics[width=\linewidth]{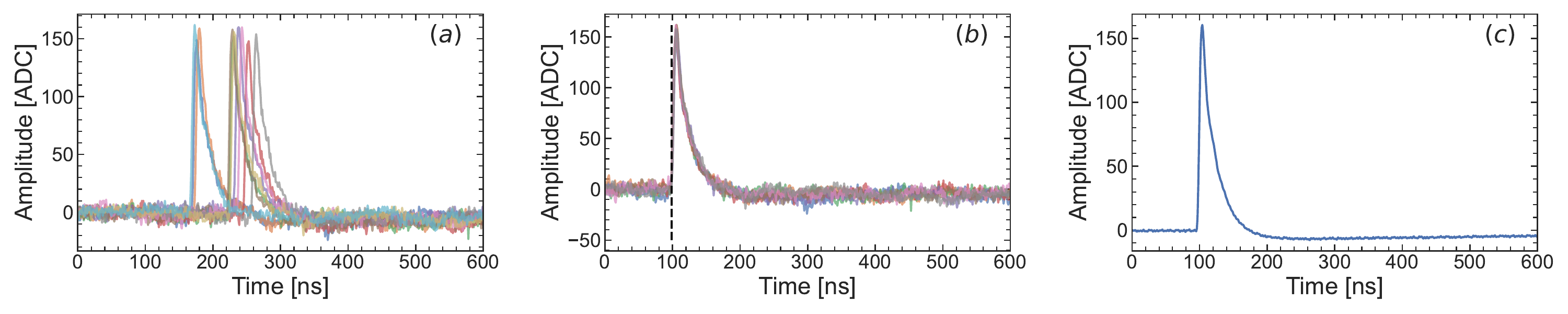}
    \caption{Template construction steps. (a) After removal of baseline, single photoelectron waveforms are selected to represent the impulse response of the \glspl{pmt}. (b) The rising edges of the peaks are aligned, and the average amplitude is computed for each time sample. (c) The resulting average waveform constitutes the \textit{template} of the \gls{pmt} \gls{spe} response for our setup.}
    \label{fig:template}
\end{figure}

The concept behind template computation, as presented in Ref.~\cite{Grassi_2018}, is to characterize the \gls{pmt} behavior via its mean impulse response (\autoref{fig:template}). The template is obtained through the following steps: (1) selection of \gls{spe} waveforms and subtraction of the baseline (\autoref{fig:template}a); (ii) time alignment of the waveforms (\autoref{fig:template}b); and (iii) calculation of the mean system response (\autoref{fig:template}c).

This template serves as a representative model of the signal acquired by a \gls{pmt} and is used to derive the signal \gls{psd}. The signal \gls{psd} is computed as the squared absolute value of the template's \gls{fft}; a more in-depth description of this process is provided in \autoref{app:psd}. \autoref{fig:psd}a illustrates the signal \gls{psd} for our experimental setup; a noticeable peak is present at around 250~MHz: this is to be attributed to a distortion caused by the detector emulator used in our experimental setup.

\subsection{Noise characterization}
\begin{figure}[htbp]
    \includegraphics[width=\linewidth]{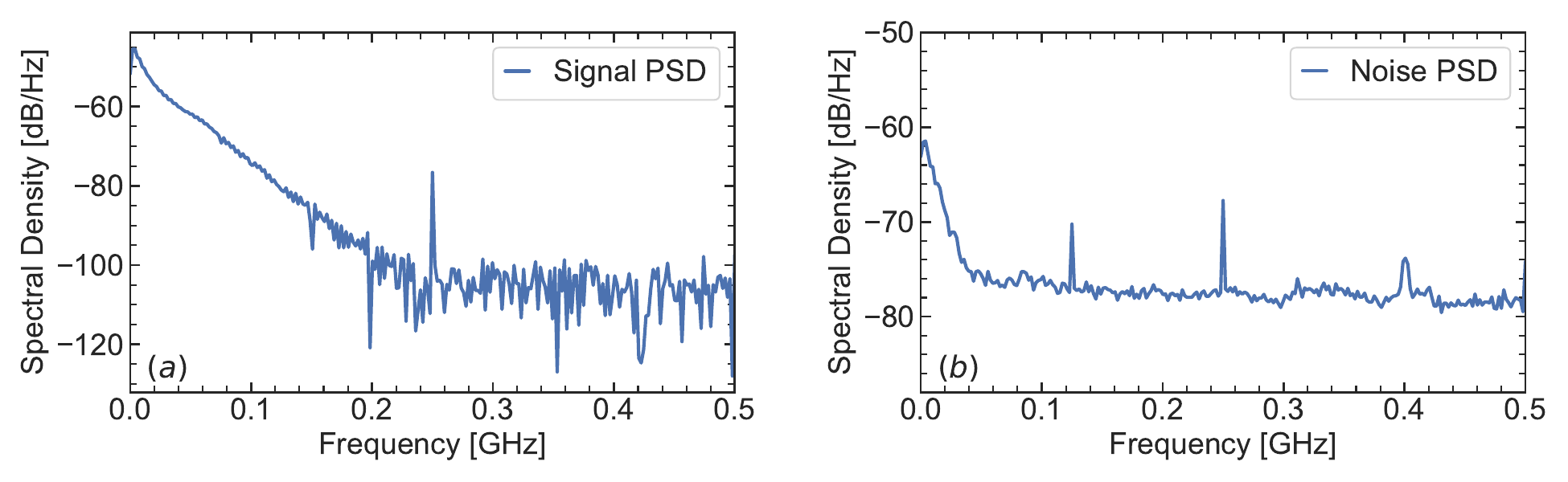}
    \caption{Power Spectral Density characterization. (a) The Power Spectral Density of the signal template shown in \autoref{fig:template}. (b) The noise Power Spectral Density, featuring both white noise (constant across the frequency domain) and colored noise spikes. The peak at 250 MHz is present in both PSDs and is an artifact of environmental noise from the power grid coupling into the setup through the emulator’s power supply.}
    \label{fig:psd}
\end{figure}

A similar approach is applied to noise characterization. To better characterize the effect of the 250~MHz peak, we left the detector emulator switched on plugged in the \gls{gcu} and a series of "empty waveforms" is acquired periodically without a signal trigger, selecting only waveforms containing noise contributions from the electronics or the environment. The \gls{npsd} is estimated as the mean of the \glspl{psd} of these waveforms. The noise \gls{psd} for our experimental setup is shown in \autoref{fig:psd}b, where the 250~MHz peak is still visible. This peak is to be attributed to the detector emulator: it's an artifact of environmental noise from the power grid coupling into the setup through the emulator’s power supply.

\subsection{Frequency response and FIR filters}
\label{sec:fir}
\subsubsection{Wiener filter}
Once the signal and noise \glspl{psd} are characterized, the Wiener filter~\cite{vaseghi2008advanced} frequency ($f$) response is calculated as:
\begin{equation}
    W(f) = \frac{SNR(f)}{SNR(f)+1}, \quad \text{where} \quad SNR(f)=\frac{SPSD(f)}{NPSD(f)}.
    \label{eq:wf}
\end{equation}
The Wiener filter is approximated with a Type I \gls{fir} filter~\cite{DFD} using the \texttt{firls} method from the Python library \texttt{scipy.signal}~\cite{scipy}, which implements the \gls{lmse} design method~\cite{DFD}. The symmetric frequency response of the Type I filter makes it the optimal choice for noise rejection, as it acts as a \textit{smoothing} filter. \autoref{fig:spectcompare}a compares the frequency response of the \gls{fir} filter with the frequency response of the original filter. Although the responses appear significantly different, this is expected, as the error minimization of the \gls{lmse} method is constrained by the limited number of coefficients available.

\subsubsection{Deconvolution filter}
\begin{figure}[htbp]
    \centering
    \includegraphics[width=\linewidth]{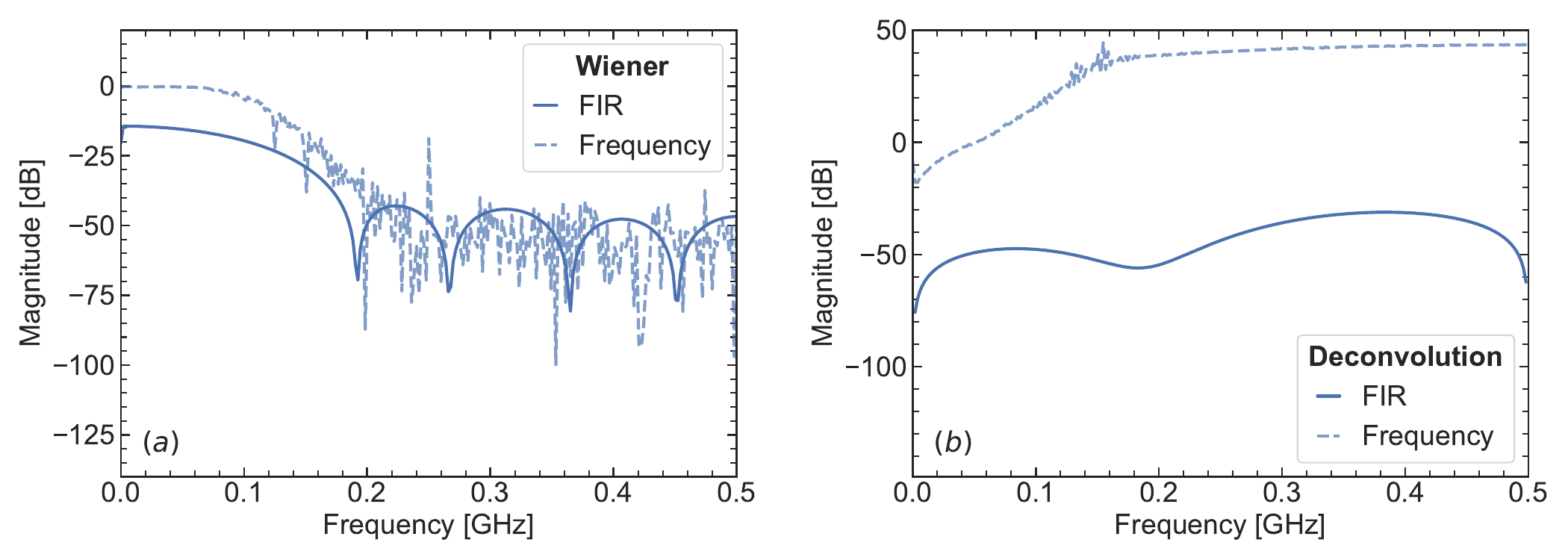}
    \caption{\gls{fir} filter frequency response comparison. The \gls{fir} filter response deviates from the ideal frequency response for both the Wiener (a) and Deconvolution (b) filters. This discrepancy results from the approximation inherent in the \gls{lmse} method when minimizing error with a limited coefficient count. The dips observed in the FIR response in panel (a) correspond to the zeros of the filter's frequency response. However, this is not an issue, as the objective is to construct filters that exhibit the desired behavior in the time domain.}
    \label{fig:spectcompare}
\end{figure}

The Deconvolution filter is computed as the reciprocal of the template's frequency amplitude spectrum $H(f)$:
\begin{equation}
    D(f)=\frac{1}{H(f)},  
\end{equation}
It is converted into a Type III \gls{fir} filter using the \texttt{remez} method from \texttt{scipy.signal}, an implementation of the Remez exchange algorithm~\cite{remez} for \gls{fir} filter design. The Type III filter acts as a \textit{differentiator}, making it ideal for deconvolution, as it enhances the rising edges of the waveform, thereby directly supporting peak detection by highlighting points of maximum derivative. \autoref{fig:spectcompare}b compares the frequency response of the \gls{fir} filter with the frequency response of the starting filter.

\glsreset{rtwd}
\section{Real-Time Wiener Deconvolution implementation in JUNO}
\label{sec:rtwd}

In this section, we present the implementation of the \gls{rtwd} algorithm within the \glsxtrshort{juno} \glsxtrshort{gcu} \glsxtrshort{fpga}, optimized to enhance the reconstruction of \glspl{tq} for \glspl{lpmt}.
% \autoref{fig:block} illustrates the block diagram of the \glspl{tq} reconstruction algorithm.
The signal processing pipeline operates sequentially: first, the waveform baseline is evaluated. Next, an over-threshold block identifies hit candidates and subtracts the baseline. The \gls{rtwd} algorithm is then applied to the baseline-subtracted signal, followed by a peak finder that reconstructs the final \gls{tq} pairs.
This section focuses on the high-level functionality of these blocks, while \autoref{app:rtl} provides a detailed description of the specific \gls{rtl} implementation.

% \begin{figure}[htbp]
%     \centering
%     \includegraphics[width=0.8\columnwidth]{images/block.pdf}
%     \caption{Block diagram of the TQ reconstruction algorithm. The developed algorithm is schematized as follows: first, waveforms from the Flash ADCs are processed by a Baseline Tracker module to extract baseline information; second, the Threshold Check module uses this information to invert the waveform polarity and identify potential peaks; third, the waveforms are filtered by the FIR filter modules; finally, the Peak Finder module identifies peaks produced by photoelectron hits.}
%     \label{fig:block}
% \end{figure}

\subsection{Baseline Tracker}
A positive voltage supply was selected for the \glspl{pmt} because the photocathode is in contact with the water; consequently, when a photon strikes the \gls{pmt}, the resulting waveform shows a negative spike relative to the baseline, and a small fluctuation with  opposite polarity called \textit{overshoot} \cite{Liu_2023}. To fully capture the whole signal, namely to accomodate both polarities in the 14-bit range of the Flash \glspl{adc}, an offset is applied during waveform digitization. 
%to represent the voltage amplitude including the overshoot, a fluctuation of the baseline below its average value\cite{Liu_2023}. 
A direct consequence of this configuration is the necessity to characterize the baseline of the Flash \gls{adc} output 
%(referred to as \textit{channels}) 
to monitor its evolution and account for baseline drift caused by both the power supply and \gls{pmt} aging.

\begin{figure}[htbp]
    \centering
    \includegraphics[width=0.9\linewidth]{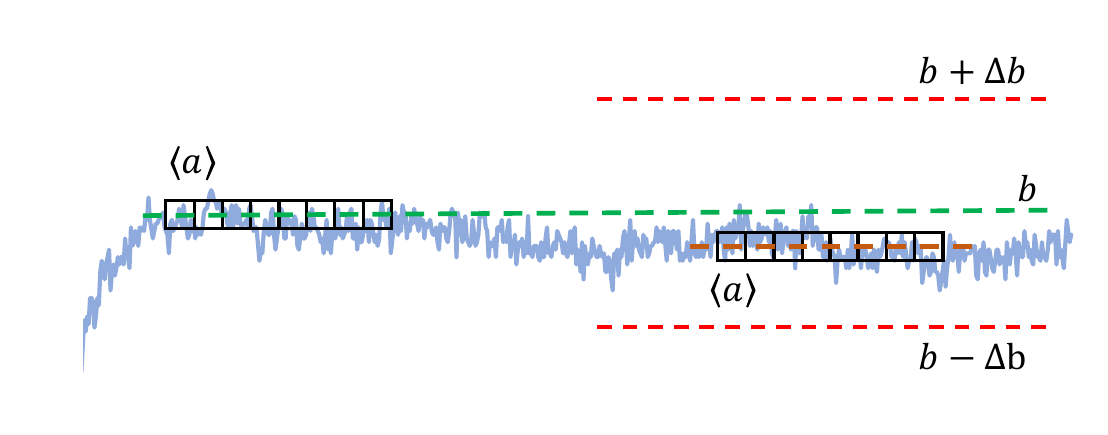}
    \caption{Baseline tracking operations. The baseline tracking module identifies instabilities in the baseline value. Since the arrival time of a peak is unknown, groups of eight samples are averaged and compared with the current baseline value $b$. If the new value is compatible with normal fluctuations, the baseline is updated. To prevent a peak with a slow rising slope from causing the baseline to drift downward undetected, the updated baseline is calculated as the average of the old baseline and the current sample batch.}
    \label{fig:base}
\end{figure}

The Baseline Tracker is designed to follow baseline variations while ignoring fast transients caused by \glspl{pe}. As shown in \autoref{fig:base}, at the beginning of the acquisition, the baseline is initialized to the average of the amplitude $a_i$ of the first eight waveform samples. As each subsequent packet of eight samples arrives from the \gls{serdes}, its average $\langle a\rangle$ is calculated. The algorithm then determines whether this value is consistent with a baseline fluctuation or indicates the presence of a \gls{pe}. If $\langle a\rangle$ falls within the interval around the previous baseline $b$ defined by the red dotted lines in \autoref{fig:base}, the baseline is updated to the mean of $b$ and $\langle a\rangle$. The delta variation parameter $\Delta b$, which defines this interval, can be tuned to track the baseline accurately; currently it is set to 3 \gls{adc} counts, which matches the expected average noise level of a \gls{gcu} channel~\cite{liu2025juno20inchpmtelectronics}. To prevent the baseline from latching onto an incorrect value due to poor initialization (e.g., if the first eight samples coincide with a \gls{pe} hit) or other anomalies, the baseline is reset every 1000 samples to the average of the current eight samples.

Once the baseline has been found, it is subtracted from each waveform sample so that when there is no activity the sample amplitude fluctuates around zero. At the same time we invert the waveform polarity, in order to have the PE-induced spikes to be positive. 

\subsection{Over-threshold Check}
% Since the PMT charge in negative, 

% The waveform has negative polarity, as a result of the PMT charge being negative. We subtract its baseline, so that 

% The baseline subtraction is achieved through the operation $b-$
% value ($b$) is subtracted from each waveform sample 
% Once the baseline is established, it is subtracted from the sample values to invert the waveform polarity. 

% Then 

Each waveform sample is compared to a threshold. If a sample exceeds this threshold, a trigger primitive is generated, and the algorithm transitions to the charge reconstruction state. The algorithm continues to reconstruct the charge as long as the samples remain above the threshold, after which it returns to an idle state. The trigger signals generated by this block are also utilized by the \gls{ctu} to identify events that require the storage of \glspl{wf}.

\subsection{FIR filters}
\begin{figure}[htbp]
    \centering
    \includegraphics[width=\linewidth]{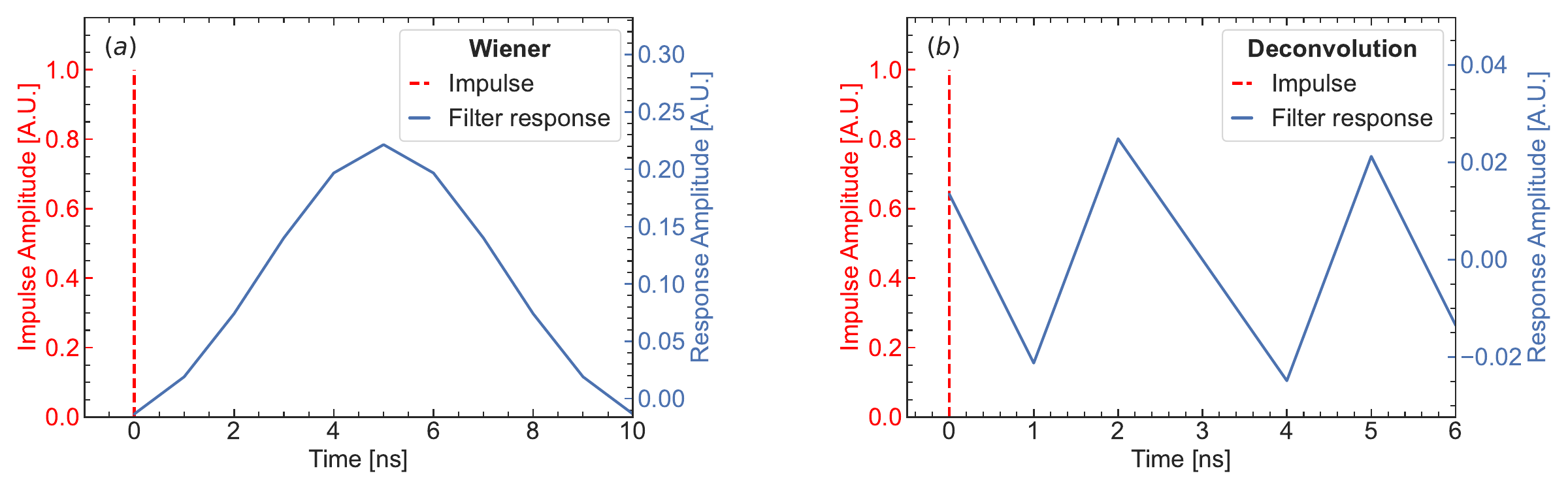}
    \caption{Impulse responses of the \gls{fir} filters. The figure illustrates the response of the designed \gls{fir} filters to a unitary impulse. The FIR Wiener filter \textit{(a)} exhibits a Type I (symmetric) 11-tap impulse response, while the FIR Deconvolution filter \textit{(b)} exhibits a Type III (antisymmetric) 7-tap response, also known as a \textit{differentiator}.}
    \label{fig:firimp}
\end{figure}

As discussed in \autoref{sec:fir}, to properly reconstruct the \gls{tq} pairs, we implemented two filters: a Type I 11-tap FIR filter approximating the Wiener filter component of the algorithm (\autoref{fig:firimp}a), and a Type III 7-tap FIR filter approximating the \gls{ifd} (\autoref{fig:firimp}b). 
Each filter acts in the time domain as a sliding window, where  \textit{taps} are coefficients multiplying a continuous stream of waveform samples produced by the FADC.
%\gls{fadc}.  
%
\autoref{fig:firimp} shows the output of both filters (blue curves) when processing an instantaneous unitary pulse (red curve). Since at each clock cycle the filter is applied to a waveform where only one sample is different from zero, each value of the response amplitude (the output of the filter) is numerically equivalent to the taps.

\begin{figure}[htbp]
    \centering
    \includegraphics[width=0.9\linewidth]{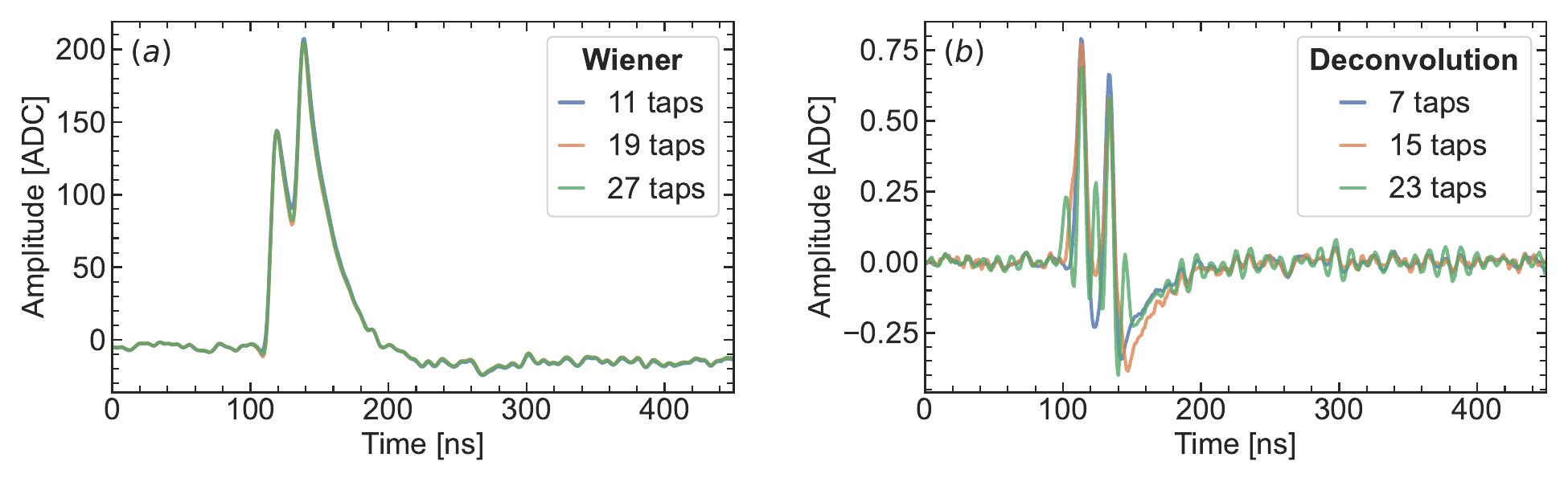}
    \caption{Effect of increasing \gls{fir} filter length.For the Wiener filter (a), increasing the length by a factor of four yields identical results. For the Deconvolution filter (b), increasing the length degrades performance. This behavior stems from the nature of the filter responses: while adding taps to a Type I filter primarily increases output delay, adding taps to a Type III filter increases sensitivity to noise spikes and slope variations. Note that in both cases, the delay introduced by the filter has been corrected to facilitate comparison.}
    \label{fig:fir-compare}
\end{figure}

%The \textit{taps} shown in \autoref{fig:firimp} are coefficients multiplying waveform samples over a sliding window
Although the filter lengths are short compared to the readout window ($1008$ ns in the default 
\gls{juno} configuration), increasing the number of coefficients does not improve performance, as demonstrated in \autoref{fig:fir-compare}a. In the case of deconvolution, extending the filter actually degrades the results (\autoref{fig:fir-compare}b).

Both filters are designed prior to algorithm execution based on a dataset of approximately 10,000 readout windows. They are implemented using the Vivado FIR Compiler IP, which allows efficient integration of FIR filters into the HDL design, although with reduced low-level control. The Vivado FIR Compiler IP supports dynamic coefficient reloading, enabling modification of the filter's impulse response. This is managed via the \gls{gcu} slow control using the IPbus protocol~\cite{ipbus}\cite{ipbus-juno}. A set of new coefficients can be loaded into a FIFO memory within the \glspl{tq} reconstruction block via a series of instructions and used to reconfigure the filters. This operation, which must be performed at every \gls{gcu} start-up, is automated by a set of custom scripts.

\subsection{Peak finding}
\begin{figure}[htbp]
    \centering
    \includegraphics[width=0.6\columnwidth]{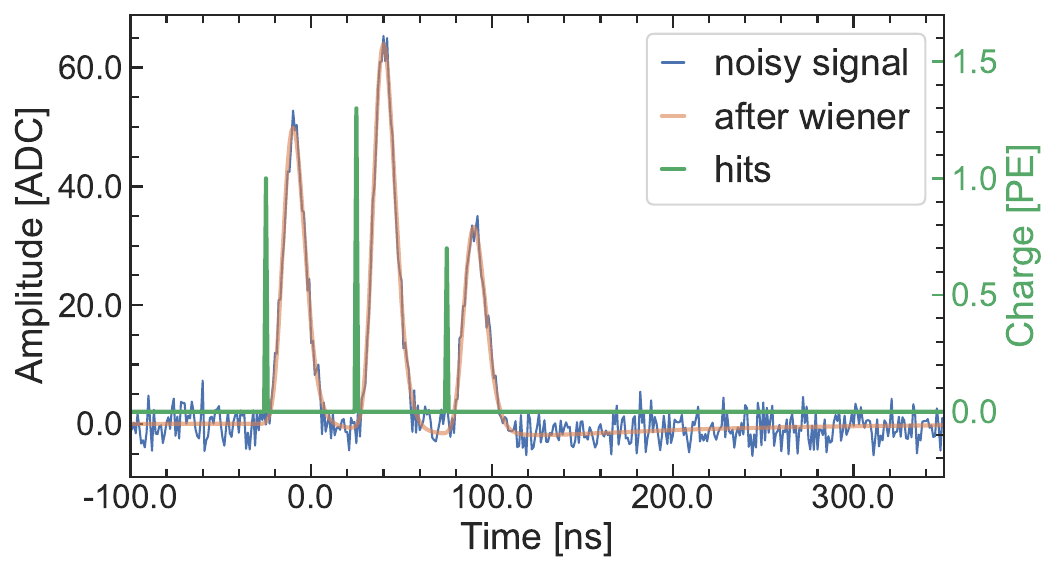}
    \caption{Ideal Wiener Deconvolution Result. In an ideal scenario, the Wiener filter maximizes the \gls{snr} by eliminating noise (orange), while the Deconvolution removes the acquisition system (\glspl{pmt}) response (green). This produces an output of spikes that perfectly represent the input \gls{pe} charge. The waveforms shown here are simulated.}
    \label{fig:w_d}
\end{figure}

The final stage of the \gls{rtwd} algorithm is the identification of peaks in the FIR Deconvolution filter output. Ideally, after deconvolution, the waveforms consist solely of \gls{pe} spikes: \autoref{fig:w_d} shows a simulated waveform with three \glspl{pe} hits (blue), the waveform after the application of the ideal Wiener filter (orange), and the final result of the ideal Deconvolution (green) that corresponds exactly to the input charge received by the \gls{pmt}.

Correct identification of these spikes and evaluation of their amplitude ensures an accurate \gls{tq} reconstruction.

\begin{figure}[htbp]
    \centering
    \includegraphics[width=0.65\columnwidth]{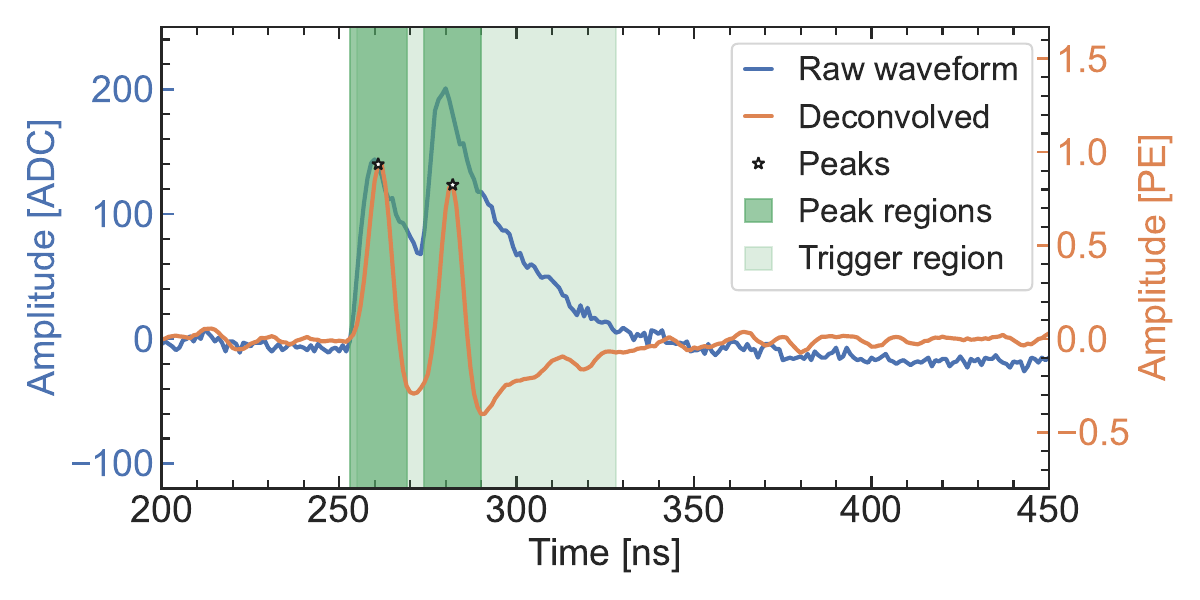}
    \caption{Peak identification on deconvolved waveform. The figure demonstrates the peak identification process. The trigger region (light green) is defined as the interval where the raw waveform (blue) exceeds the threshold. The peak regions (dark green) are established around peaks identified within the trigger region on the deconvolved waveform (orange).}
    \label{fig:peak}
\end{figure}

To achieve this, a coarse peak identification is first performed on the "raw" waveform. Subsequently, after both filters are applied, the precise peak is located within the time window indicated by the coarse identification. This process ensures that a search is initiated whenever the waveform exceeds the threshold, but a charge value is output only if the detected peak resembles a \gls{spe}. This is verified by confirming that the peak is the maximum value within a programmable interval (\autoref{fig:peak}).

The peak value corresponds exactly with the charge of the \gls{pe} and the hit time is identified as the timestamp corresponding to the peak sample. The \glsxtrlong{tq} is encapsulated in a 128-bit packet and sent to the \gls{daq}.

\glsreset{rtwd}
\section{Results}
\label{sec:result}
The \gls{rtwd} algorithm was validated using the experimental setup shown in \autoref{fig:exp_setup}. The use of a detector emulator allowed us to test its performance under various configurations of hit amplitude and temporal separation, comparing it against other reconstruction algorithms, specifically \gls{coti} and offline Deconvolution (described later in this section).

We studied four values (0.030, 0.037, 0.050, 0.076) of the ratio between the noise \glsxtrshort{rms} ($\alpha$) and the mean  peak amplitude ($\beta$) of a hit, to cover a wide range of acquisition scenarios. Although all configurations yielded compatible results, we focused on the ratio $\frac{\alpha}{\beta}=0.037$, as this scenario best approximates the standard \gls{juno} operating conditions. For each noise ratio, we acquired data using hit pairs with increasing temporal separations $\Delta t$. 
The investigated $\Delta t$ values normalized to the mean peak width were 0.5, 0.7, 1.0, and 1.2.
This range was chosen to explore conditions varying from near signal superposition (0.5) to clearly distinguishable hits (1.2). Finally, a control dataset of \gls{spe} waveforms was acquired for each noise level.

\begin{figure}[htbp]
    \centering
    \includegraphics[width=\linewidth]{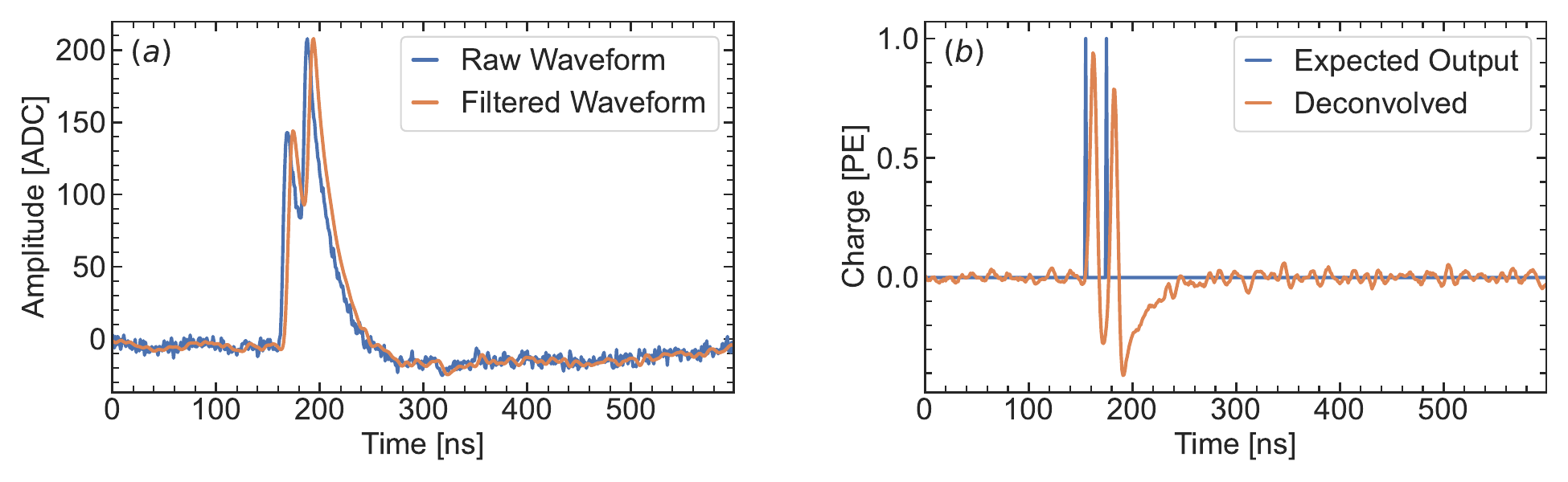}
    \caption{\glsxtrlong{rtwd} Results.
    (a) A waveform comprising two hits before (blue) and after (orange) the application of the Weiner filter. As expected, this filter increases the \gls{snr} without distorting the general waveform shape. A delay introduced by the filter is visible, which is characteristic of the FIR implementation.
    (b) The Deconvolution filter produces a waveform featuring  sharp peaks, comparable to the ideal result (expected output) in amplitude and width. The deconvolved  waveform exhibits a slightly reduced peak amplitude due to residual noise and a prominent undershoot. While not problematic for single \gls{pe} hits, this undershoot can degrade charge estimation for closely spaced hits, because a subsequent hit arriving in close proximity suffers from charge underestimation. The magnitude of this effect depends on the temporal distance between the two hits and charge of the first hit.}
    \label{fig:result}
\end{figure}

As shown in \autoref{fig:result}, the algorithm yields good results by correctly identifying two separate hits and by yielding a reasonable estimate of the hit charge. However, it struggles to completely remove the effects of the  \gls{pmt} response.
Panel (a) of \autoref{fig:result} shows the capability of the Wiener filter to significantly enhance the \gls{snr} without distorting the waveform.
%
%Examining the effect of the Wiener filter in \autoref{fig:result}a, reveals that the filter significantly enhances the \gls{snr} without distorting the waveform:
Namely, the filtered waveform (orange) displays a lower noise content than the raw waveform (blue), particularly in the baseline region. 
Panel (b) shows the output of the Deconvolution filter, 
%is shown in \autoref{fig:result}b
 where the amplitude of both peaks in the deconvolved waveform is compatible with the expected unitary input charge, although a sizable overshoot is still present.
%
% If we apply the Deconvolution filter to the waveform, we can see that the first deconvolved peak amplitude is comparable to the input hit charge (i.e., the expected result), but the waveform in \autoref{fig:result}b remains characterized by the presence of significant undershoots.
%
This makes the reconstruction of subsequent hits challenging,
%: as shown in the figure, the charge 
because the charge of the second hit is underestimated 
when the delay between the two hits is
%the second \gls{pe} arrives with a delay 
shorter than the width of the \gls{spe} template. The maximum underestimation is approximately one-third of the \gls{spe} amplitude for a pair of \glspl{pe}, and it may be larger if multiple \glspl{pe} strike the \gls{pmt} in rapid succession. This effect can be calibrated out.

%%%%%%%%%%%%%%

The primary improvement in \gls{tq} reconstruction offered by FIR Deconvolution is its ability to correctly identify the number of \gls{pe} hits when they  occur in close temporal proximity.
To compare \gls{rtwd} and \gls{coti} performance, both reconstruction algorithms are executed over the  datasets mentioned above, where the $\Delta t$ among two hits is kept constant within each dataset. An additional dataset featuring SPE waveforms is used as a control sample, because both methods are expected to perform similarly. 

To evaluate the asymptotic best performance of the deconvolution method, we processed all the data also with a deconvolution algorithm running offline, that is, on a computer rather than on the FPGA. This algorithm is implemented in the frequency domain, similarly what JUNO does in its official reconstruction \cite{Zhang_2019}. As already mentioned, the frequency-based deconvolution cannot be implemented in the FPGA, but it provides us with a  benchmark to assess the \gls{rtwd} performance.
%Datasets are also processed offline with an implementation of the Deconvolution in the frequency domain, to check the performance of the online algorithms compared to the standard algorithm used by \gls{juno} to reconstruct \glspl{tq} offline \cite{Zhang_2019}.
 
\begin{figure}[htbp]
    \centering
    \includegraphics[width=0.75\linewidth]{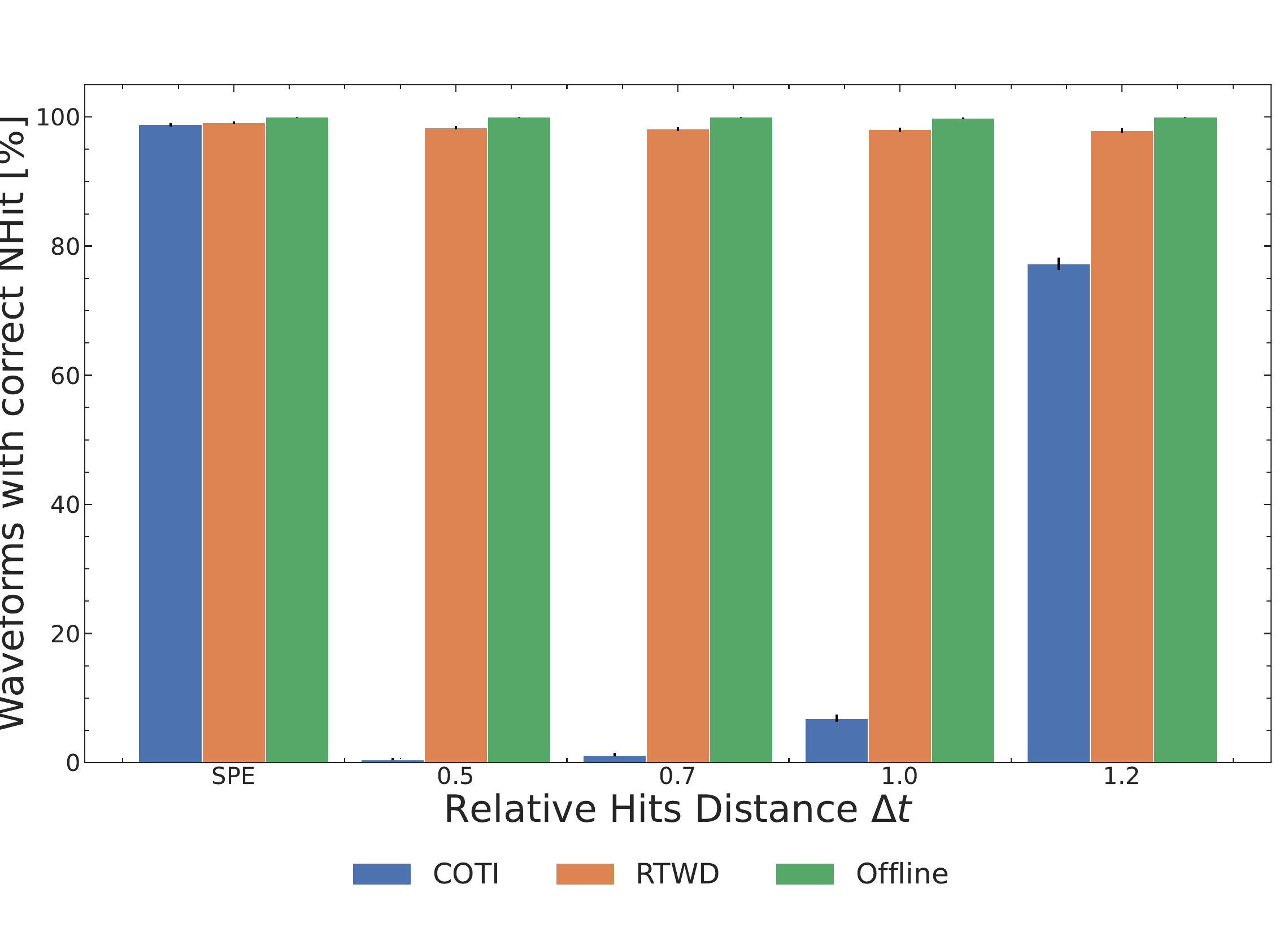}
    \caption{Evaluation of the peak identification performance. We test three algorithms and we compare the fraction of events where the number of identified peaks is correct. 
    %   
    % To properly compare the algorithms, it is necessary to quantify their ability 
    % to identify the correct number of peaks. 
    Offline reconstruction (green) successfully reconstructs peaks in all scenarios; \gls{rtwd} (orange) demonstrates performance comparable to offline reconstruction; \gls{coti} (blue) exhibits generally inferior performance compared to the other methods. While \gls{rtwd} runs on the FPGA, i.e., with limited hardware resources, \gls{coti} and Offline run on CPU with virtually unlimited computational resources.}
    \label{fig:nhit}
\end{figure}

\autoref{fig:nhit} shows the fraction of waveforms where the number of hits was correctly reconstructed, as a function of the relative peak distance among the hits; the performance of FIR Deconvolution (RTWD) is consistently comparable to Offline Deconvolution. In contrast, \gls{coti} begins to resolve the two peaks only when the relative distance exceeds 1.0.

\begin{figure}[htbp]
    \centering
    \includegraphics[width=0.95\columnwidth]{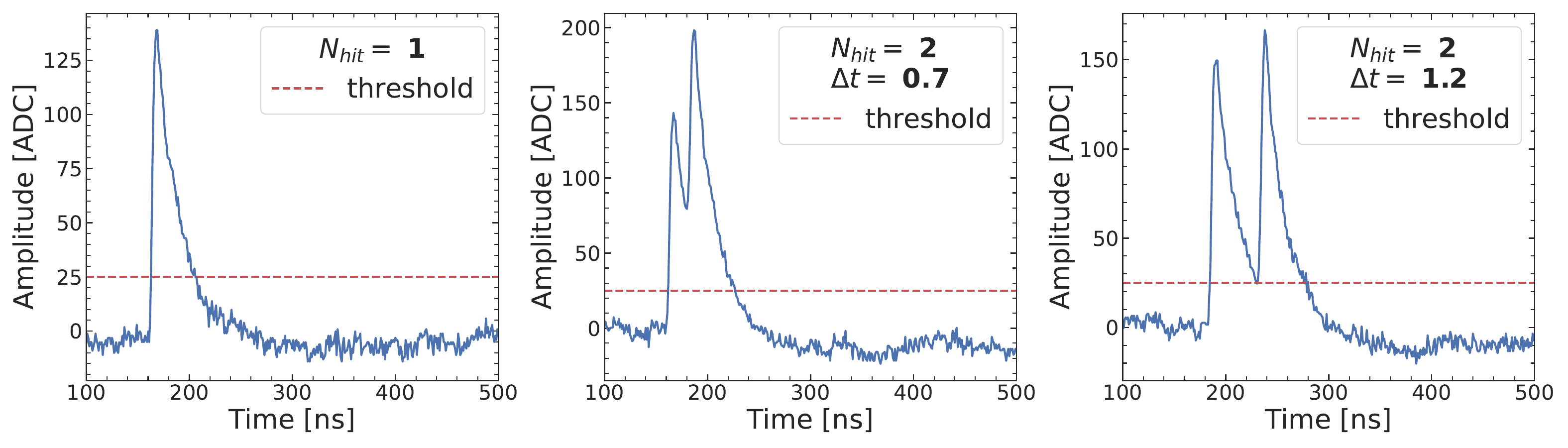}
    \caption{Waveforms for closely spaced hits at varying distances. The threshold, shown in red, is never sufficiently high to allow \gls{coti} to resolve the two peaks. Increasing the threshold would result in charge underestimation.}
    \label{fig:wfs}
\end{figure}

The power of the \gls{rtwd} algorithm lies in properly accounting for the width of each peak when reconstructing the number of hits and  the total charge. This can be effectively visualized using waveforms whose amplitude does not fall below the integration threshold between the first and the second hit, therefore preventing  \gls{coti} from integrating the two peaks separately, as shown in  \autoref{fig:wfs}. 

%%%%%%%%

% The limitation of
% This limitation of \gls{coti} is further illustrated in \autoref{fig:nhit-perc} and \autoref{fig:wfs}. To ensure proper charge estimation, the threshold is set at a level that prevents the \gls{coti} algorithm from distinguishing closely spaced peaks; the waveform does not return below the threshold until the peaks are sufficiently separated, resulting in a high percentage of events where two peaks are reconstructed as a single hit.

To better characterize the behavior of the two algorithms, we process the datasets with two true hits and we report in \autoref{tab:nhit-perc} the fraction of waveforms where the reconstructed number of hits is larger or smaller than 2.
%
%For each two-hit dataset the percentage of waveforms reconstructed more than two hits and the percentage of waveforms reconstructed with less than two hits.  
%
% Apart from the SPE dataset, all the others have been generated by injecting two separate peaks into the readout electronics. In each bar, the dark area represents the fraction of waveforms where the number of hits was correctly identified and is equivalent to what is shown in \autoref{fig:nhit}.
\begin{table}[]
    \centering
    \begin{tabular}{lcccccccc}
    \hline
    $\mathbf{\Delta t}$ & \multicolumn{2}{c}{\textbf{0.5}} & \multicolumn{2}{c}{\textbf{0.7}} & \multicolumn{2}{c}{\textbf{1.0}} & \multicolumn{2}{c}{\textbf{1.2}} \\
    \textbf{Reco} $\mathbf{N_{hit}}$             & $< 2$           & $> 2$          & $< 2$           & $> 2$          & $< 2$           & $> 2$          & $< 2$           & $> 2$          \\ \hline
    \textbf{Offline}    & \multicolumn{2}{c}{-}            & \multicolumn{2}{c}{-}            & 0.2\%           & -              & \multicolumn{2}{c}{-}            \\
    \textbf{COTI}       & 99.5\%          & -              & 98.8\%          & -              & 93.2\%          & -              & 18.8\%          & 3.9\%          \\
    \textbf{RTWD}       & 1.6\%           & -              & 1.9\%           & -              & 1.9\%           & -              & 2\%             & -             
    \end{tabular}
    \caption{Percentage of wrongly reconstructed waveforms. The table shows how many waveforms each algorithm wrongly reconstructs, specifying if it reconstructs less or more hits than the two actually presents in the waveform. While RTWD shows a general capacity of properly reconstruct the correct number of hits for all the value of $\Delta t$, COTI struggles to identify the two peaks until they are separated by a considerable time distance.}
    \label{tab:nhit-perc}
\end{table}
\gls{rtwd} shows consistent performance for all datasets,
while the values reported in the table confirm
%This table confirms the fact 
that the main issue of \gls{coti} is its inability to distinguish close hits, merging them into a single reconstructed \gls{tq} pair.
% \begin{figure}[htbp]
%     \centering
%     \includegraphics[width=0.75\linewidth]{images/Nhit_percentage.pdf}
%     \caption{Analysis of peak identification efficiency. The figure illustrates how the inability of \gls{coti} to identify peaks leads to an overall lower hit count until the peaks are sufficiently separated. This suggests that for closely spaced peaks, the waveform does not return below the threshold (further explained in \autoref{fig:wfs}).}
%     \label{fig:nhit-perc}
% \end{figure}

\begin{figure}[htbp]
    \centering
    \includegraphics[width=0.75\columnwidth]{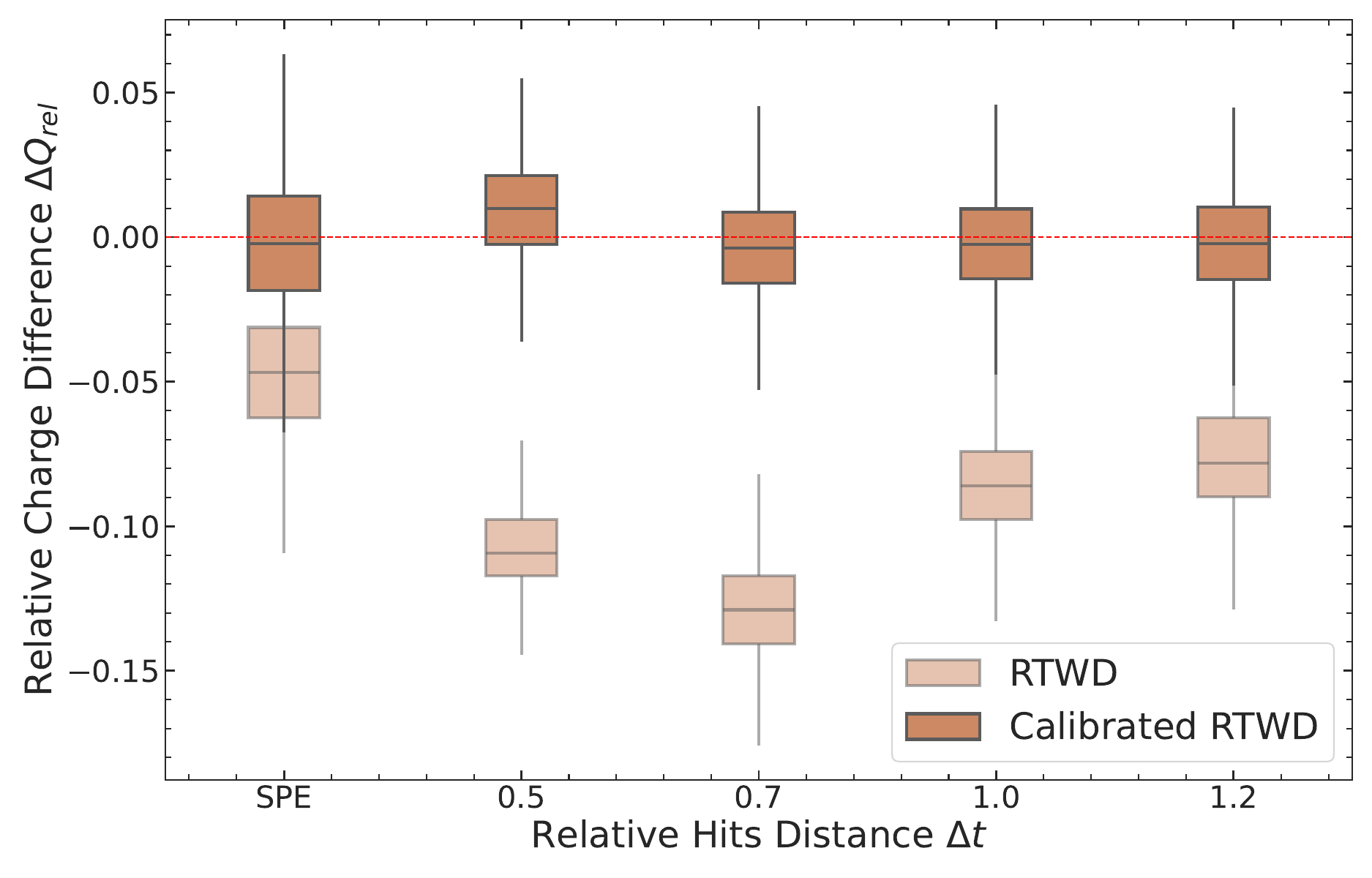}
    \caption{Effects of undershoot and residual noise. As shown, residual noise generates a relative negative bias in charge estimation of $\sim5\%$ for SPE dataset. The figure also demonstrates how the temporal distance between two peaks impacts the charge estimation. Both effects can be quantified and calibrated to improve overall charge reconstruction accuracy.}
    \label{fig:violin}
\end{figure}

The significant difference in the hit detection efficiency offers an opportunity to improve the reconstruction of the waveform charge. Precise PE detection provided by RTWD allows 
to compensate 
%for the compensation of 
%
the charge misreconstruction due to the 
undershoot  and to other artifacts arising from closely spaced \glspl{pe}. These effects are not constant as they depend primarily on the relative distance between two consecutive hits, as shown by the light-colored bars in \autoref{fig:violin}.
In this figure, the horizontal axis displays the temporal distance between \gls{pe} pairs relative to the width of the template, while the vertical axis represents the distribution of the  reconstructed charge by the online algorithm  relative to the  charge reconstructed offline. As the \glspl{pe} become closer in time, the underestimation increases. 

%The aforementioned light-colored bars  illustrate the increasing impact of the undershoot on charge estimation. The x-axis displays the distance between \gls{pe} pairs relative to the width of the template, while the y-axis represents the distribution of the reconstructed charge in the online environment relative to the reconstructed charge offline. As the \glspl{pe} become closer in time, the underestimation increases. 

\begin{figure}[htbp]
    \centering
    \includegraphics[width=0.75\linewidth]{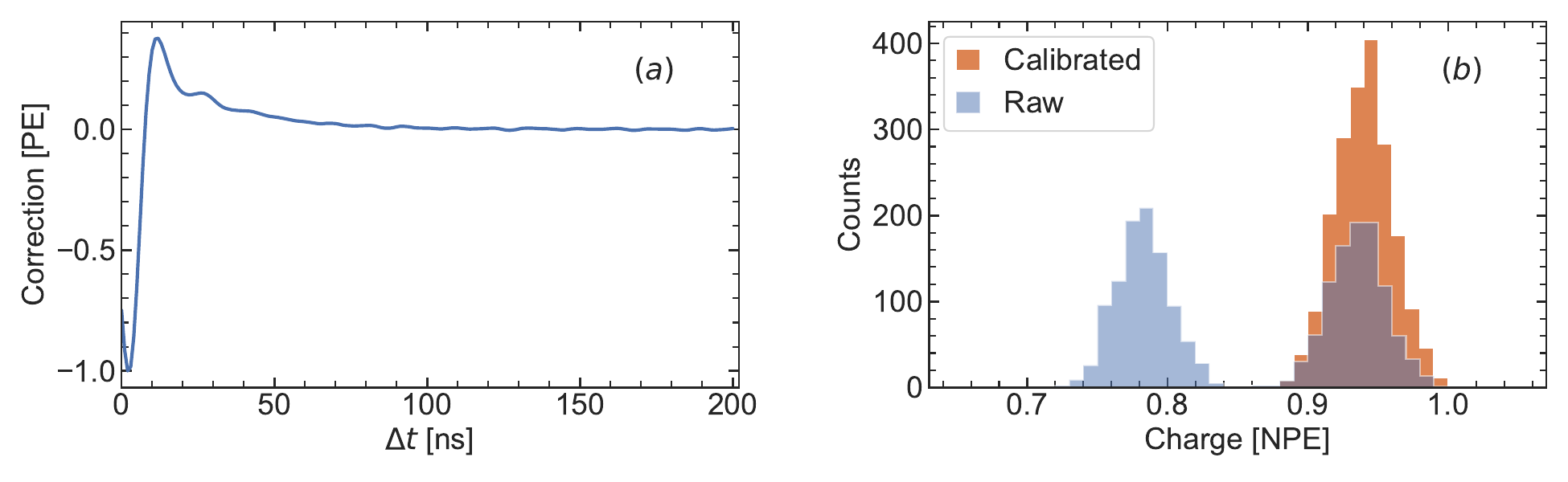}
    \caption{Calibration of reconstructed \gls{pe} charge. (a) The calibration curve, obtained from the reversed deconvolved template waveform starting from the peak value.
    (b) Before calibration, the Raw charge distribution (light blue) shows a double population: one for the first hit and one for the second, underestimated hit. After calibration, the charge distribution (orange) is restored to a single population. An additional calibration factor must be applied to all  reconstructed charges to have the mean value of the distribution matching 1 \gls{pe} (further details in the text).
%    
%    the The distribution is then shifted in order to 
%    centered on the expected mean. The residual bias of the mean value with respect to the expected value 
    }
    \label{fig:calib}
\end{figure}

A second observed effect is an overall $\sim5\%$ negative bias in the reconstructed charge. This is related to the non-ideality of the applied Wiener filter, and the resulting residual noise in the waveform. Namely, since the FIR Wiener filter approximates the ideal Wiener filter, it cannot fully maximize the waveform \gls{snr}, resulting in a systematic underestimation of the \gls{pe} charge. Both effects are quantifiable and can be calibrated offline. The overall bias can be corrected with a simple constant, estimated from the mean value of the \gls{spe} charge reconstruction (the first light-colored bar in 
%box plot in the uncalibrated charge plot of 
\autoref{fig:violin}). The undershoot correction is derived from the deconvolved waveform of the \gls{spe} template. By isolating the waveform segment following the peak, we can use it as a calibration curve for all hits that arrive at a distance $\Delta t$ from the primary peak (\autoref{fig:calib}a). The application of this calibration to a dataset comprising waveforms with two hits is shown in \autoref{fig:calib}b. 
Before calibration, the charge distribution exhibits two populations: one for the first hit (higher charge) and one for the second hit (underestimated charge). After calibration, the distribution collapses into a single population centered on the expected mean value.
The performance of the calibration procedure applied to the datasets featuring increasing $\Delta t$ values is shown by the dark bars in \autoref{fig:violin}. 

\begin{figure}[htbp]
    \centering
    \includegraphics[width=0.75\columnwidth]{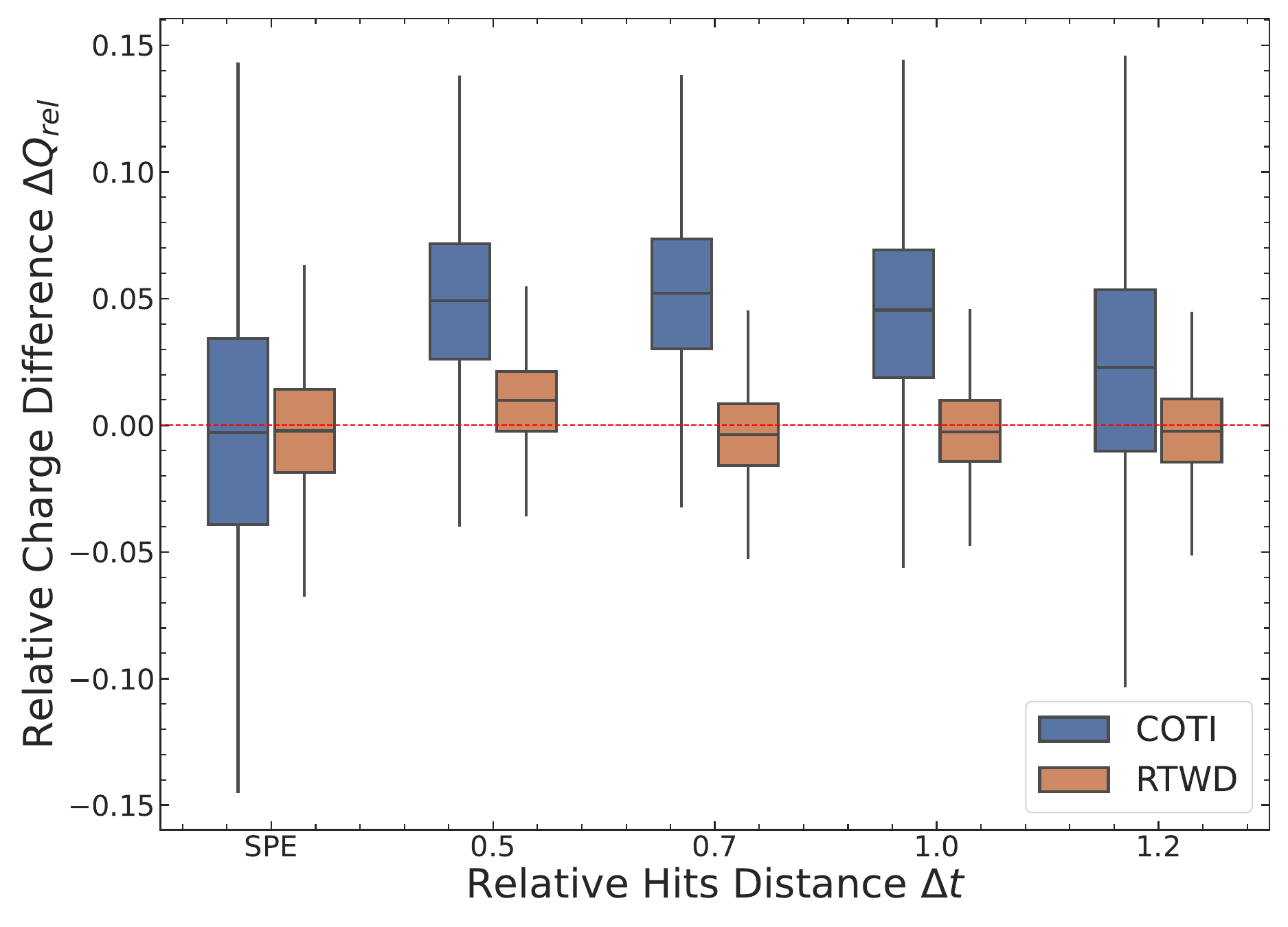}
    \caption{Comparison between \gls{coti} and \gls{rtwd}. While the relative charge reconstructed by \gls{coti} and \gls{rtwd} shows general agreement, the distributions for a relative distance minor than 1.2 highlights the advantage of \gls{rtwd}. The calibration improves the results for \gls{rtwd} relative to \gls{coti}, as the former's superior peak identification facilitates effective undershoot correction.}
    \label{fig:violin-comp}
\end{figure}

Comparing the calibrated \gls{rtwd} charge to the total charge reconstructed by \gls{coti}, regardless of the number of reconstructed hits, we observe decent agreement (\autoref{fig:violin-comp}). 
%
%However, \gls{rtwd} is significantly less affected by the presence of closely spaced \glspl{pe}.
%
However, \gls{rtwd} 
charge reconstruction is significantly more accurate because it is less affected by the presence of closely spaced \glspl{pe}. This is a consequence of 
the \gls{rtwd} algorithm's ability to distinguish close peaks, hence enabling separate calibration of the first and the second hit.
%enables effective calibration, thereby improving charge reconstruction accuracy.

\begin{figure}[htbp]
    \centering
    \includegraphics[width=0.75\columnwidth]{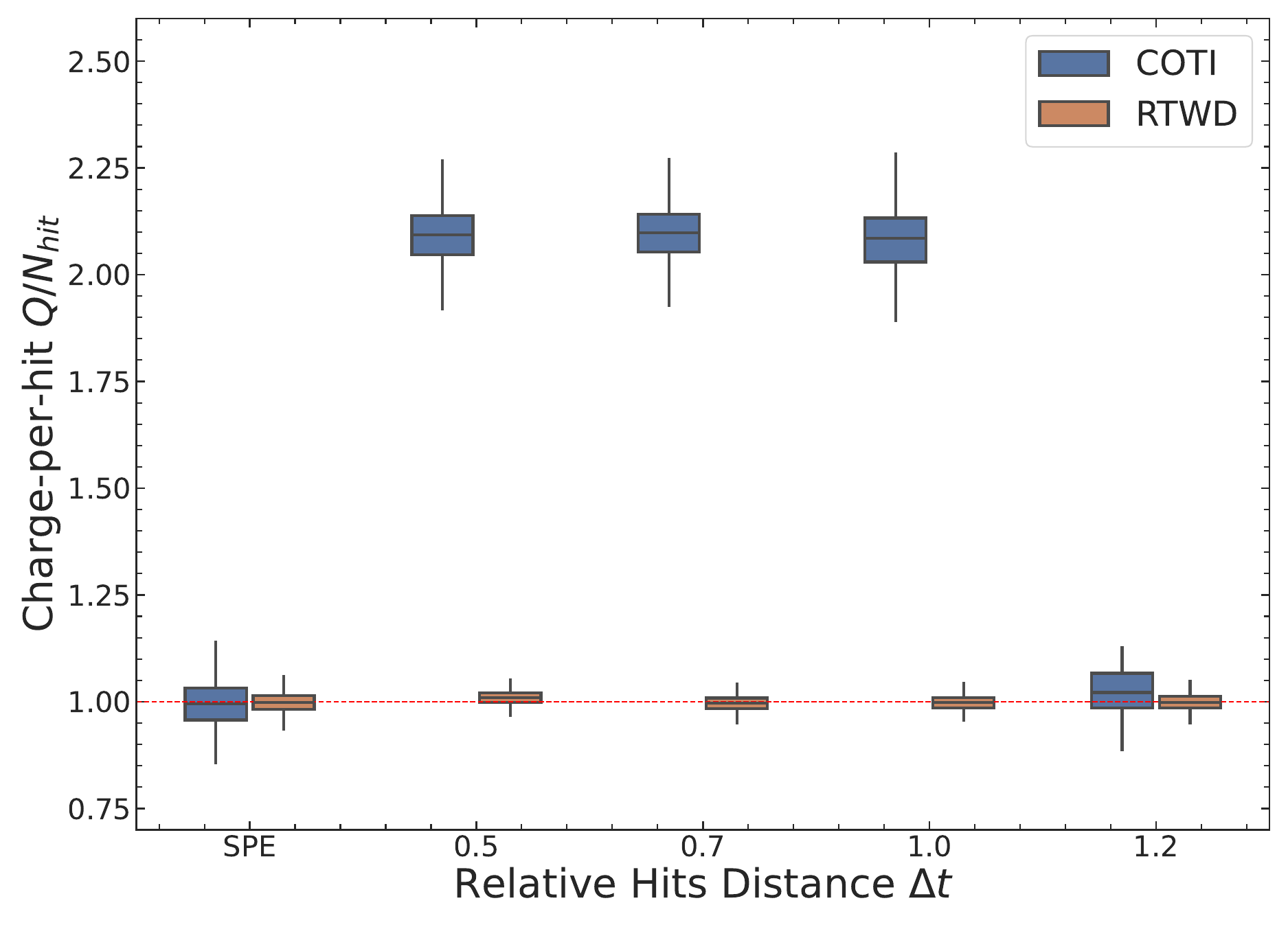}
    \caption{Comparison between \gls{coti} and \gls{rtwd}. While the relative charge reconstructed by \gls{coti} and \gls{rtwd} shows general agreement, the distributions for a relative distance minor than 1.2 highlights the advantage of \gls{rtwd}. The calibration improves the results for \gls{rtwd} relative to \gls{coti}, as the former's superior peak identification facilitates effective undershoot correction.}
    \label{fig:violin-nhit-compare}
\end{figure}

A better metric to evaluate how each algorithm performs is the reconstructed \textit{charge-per-hit} $Q/N_{hit}$. Since we inject hits of unitary charge (1 PE), if the reconstructed charge-per-hit is close to one, the algorithm is well handling the closely spaced hits and the \gls{tq} pairs generated are accurate; otherwise, the algorithm is missing one of the two peaks or merging them into a single one.
%, producing a different charge-per-hit value. 
%
\autoref{fig:violin-nhit-compare} shows how \gls{rtwd} yields a ratio close to one for each $\Delta t$, while  \gls{coti} jumps to  values close to two as a result of close peaks being misreconstructed.

\section{Conclusions}
\label{sec:conclusion}
This work presents a novel implementation of a feature extraction algorithm based on Wiener Deconvolution and \gls{fir} filters. The primary focus of the study was its application to the online reconstruction of Time-Charge pairs in the PMT of the \gls{juno} experiment. The results demonstrate a significant improvement in both peak identification and charge reconstruction in comparison to the current solution. Moreover, the performance of the \gls{rtwd} is comparable to that of offline algorithms, despite operating within the strict constraints of limited \gls{fpga} resources and the power consumption requirements of the \gls{juno} experiment.

\appendix
\glsreset{rtwd}
\section{Power Spectral Density computation}
\label{app:psd}
By definition, the \gls{psd} of a continuous function, $x(t)$, is the Fourier transform of its auto-correlation function:
\begin{equation}
    PSD = \mathcal{F}\{x(t) \circledast x^*(-t)\}=X(f)\cdot X^*(f)=|X(f)|^2.
\end{equation} 
The Fourier transform of the template $h$ in the Nyquist frequency region $-\frac{1}{2T}<f<\frac{1}{2T}$, where $T$ is the sampling period, is equivalent to the \gls{dtft}, denoted as $H_{\frac{1}{T}}(f)$:
\begin{equation}
    H(f)\triangleq H_{\frac{1}{T}}(f)=\sum^\infty_{n=-\infty}h[n]\cdot e^{-i2\pi Tfn}.
\end{equation}
However, in this specific case, we do not have access to the infinite sequence of samples required to fully describe the sampled template; we are limited to a finite subset of length $N$. Provided that $N$ is sufficiently large, the \gls{dft} $H[k]$ serves as a sampled approximation of the \gls{dtft}. Practically, the \gls{dft} is computed using the \gls{fft} numerical algorithm. In this way, we obtain a sampled version of the \gls{psd}, known as a \textit{periodogram}~\cite{McSweeney01042006}:
\begin{equation}
    PSD[k] = |FFT(h[n])|^2.   
\end{equation}

\glsreset{rtwd}
\section{VHDL Implementation Details} 
\label{app:rtl}
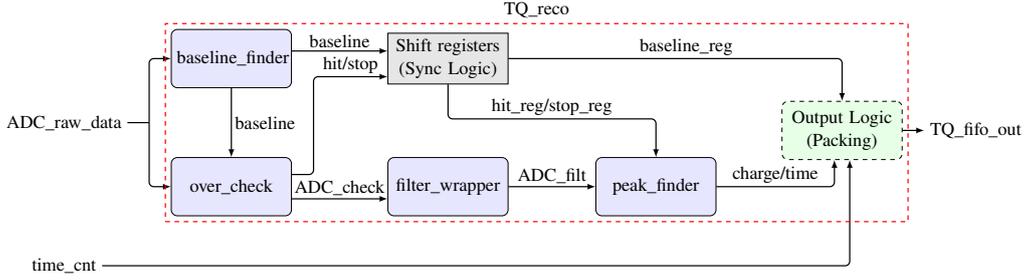
\begin{figure}[htbp]
    \centering
    \resizebox{0.9\textwidth}{!}{
    \begin{tikzpicture}[
        % Styles
        node distance=1.5cm and 1.5cm,
        top/.style={
            draw=red, 
            thick,
            dashed
        },
        block/.style={
            rectangle, 
            draw, 
            fill=blue!10, 
            text width=2.5cm, 
            align=center, 
            rounded corners, 
            minimum height=3.5em,
            font=\large
        },
        sync/.style={
            rectangle, 
            draw, 
            fill=gray!20, 
            text width=2.5cm, 
            align=center, 
            minimum height=2.5em,
            font=\large
        },
        logic/.style={
            rectangle, 
            draw, 
            fill=green!10, 
            text width=2.5cm, 
            align=center, 
            rounded corners,
            dashed,
            minimum height=3.5em,
            font=\large
        },
        line/.style={
            -latex, 
            thick,
            rounded corners=3pt
        },
        label_text/.style={
            font=\large,
            midway,
            auto,
            inner sep=2pt
        }
    ]

    % --- Nodes ---
    
    % Input Signal Name
    \node [font= \large](input) {ADC\_raw\_data};
    
    % Split point
    \coordinate [right=0.5cm of input] (split_point); 

    % Stage 1: Baseline (Top) and Over Check (Bottom)
    \node [block, above right=0.8cm and 0.5cm of split_point] (baseline) {baseline\_finder};
    \node [block, below right=0.8cm and 0.5cm of split_point] (overcheck) {over\_check};

    % Time input (Below Over Check)
    \node [below =2.7cm of input, font= \large] (time) {time\_cnt};

    % Stage 2: Sync (Top) and Filter (Bottom)
    \node [sync, right= 2.2cm of baseline] (sync) {Shift registers\\(Sync Logic)};
    \node [block, right= 2.2cm of overcheck] (filter) {filter\_wrapper};

    % Stage 3: Peak Finder
    \node [block, right=2cm of filter] (peak) {peak\_finder};

    % Stage 4: Output Logic
    \node [logic, right=of peak, yshift=1.3cm, text width=2.5cm] (logic) {Output Logic\\(Packing)};

    % Final Output
    \node [right=0.5cm of logic, font= \large] (output) {TQ\_fifo\_out};

    % --- Connections ---

    % 1. Input Distribution
    \draw [thick] (input) -- (split_point); 
    \draw [line] (split_point) |- (baseline.west);
    \draw [line] (split_point) |- (overcheck.west);

    % 2. Baseline -> Sync & Overcheck
    \draw [line] ([yshift=5pt]baseline.east) -- node[label_text] {baseline} ([yshift=5pt]sync.west);
    % Connection to Overcheck (Top Input)
    \draw [line] (baseline.south) -- node[label_text] {baseline} (overcheck.north);

    % 3. Overcheck Outputs (SEPARATED)
    % Output 1: ADC_check (Data) - Exits from UPPER East
    \draw [line] ([yshift=-8pt]overcheck.east) -- node[label_text] {ADC\_check} ([yshift=-8pt]filter.west);
    
    % Output 2: hit/stop (Control) - Exits from LOWER East
    % Goes down slightly then up to Sync
    \draw [line] ([yshift=8pt]overcheck.east) -- ++(0.5,0) |- node[label_text, near end] {hit/stop} ([yshift=5pt]sync.south west);

    % 4. Filter -> Peak
    \draw [line] (filter.east) -- node[label_text] {ADC\_filt} (peak.west);

    % 5. Sync Outputs
    % Sync to Peak (hit1/stop1)
    \draw [line] (sync.south) -- ++(0,-0.8) -| node[label_text, near start] {hit\_reg/stop\_reg} (peak.north);
    % Sync to Logic (baseline_int) - Route High
    \draw [line] (sync.east) -- ++(0.5,0) -| node[label_text, near start] {baseline\_reg} (logic.north);

    % 6. Peak -> Logic
    \draw [line] (peak.east) -| node[label_text, near start] {charge/time} ([xshift=-5pt]logic.south);

    % 7. Time Count -> Logic
    \draw [line] (time.east) -- ++(9,0) -| ([xshift=5pt]logic.south);

    % 8. Logic -> FIFO -> Output
    \draw [line] (logic.east) -- (output.west);

    \node[top,fit=(sync) (peak) (logic) (filter) (baseline) (overcheck), label={\large TQ\_reco}] (tqreco){};
    
    \end{tikzpicture}
    }

    \caption{RTL block diagram of the \texttt{TQ\_reco} entity. The raw ADC data is processed in parallel paths for baseline estimation and threshold checking, followed by filtering and peak extraction. Shift registers ensure data synchronization across the pipeline.}
    \label{fig:rtl}
\end{figure}

A scheme of the \gls{rtl} design is presented in \autoref{fig:rtl}.

% \begin{figure}[htbp]
%     \centering
%     \includegraphics[width=0.5\linewidth]{rtl.pdf}
%     \caption{Caption}
%     \label{fig:rtl}
% \end{figure}

The implementation of the \gls{tq} reconstruction logic is hosted within the top-level VHDL entity \texttt{TQ\_reco}. This module orchestrates the data path, manages the synchronization between pipelined stages, and interfaces with the system's buffering resources via FIFO memory.

\subsection{Top-Level Architecture (\texttt{TQ\_reco})}
The \texttt{TQ\_reco} entity instantiates four primary sub-modules corresponding to the algorithmic stages:
\begin{itemize}
    \item \texttt{baseline\_finder}: Estimates the signal baseline.
    \item \texttt{over\_check}: Performs dynamic thresholding and baseline subtraction.
    \item \texttt{filter\_wrapper}: Wraps the Wiener and Deconvolution \gls{fir} filters.
    \item \texttt{peak}: Extracts the charge and fine timing.
\end{itemize}

The module operates on a 128-bit wide data bus (\texttt{ADC\_raw\_data}), processing eight 16-bit samples per clock cycle. To ensure temporal consistency across the pipelined stages, shift registers (\texttt{sr\_hit}, \texttt{sr\_stop}, \texttt{sr\_baseline}) are used to delay control signals and metadata, aligning them with the processing latency of the filters and peak finder. The module also implements a Packing Output Logic to generate 128-bit \gls{tq} packets, containing information about the input channel and the characteristics of the hit.

\subsection{Baseline and Thresholding}
The \texttt{baseline\_finder} module calculates a moving average of the input signal. It utilizes an accumulator (\texttt{sum}) to compute the average of the incoming 128-bit data frame. The logic includes a stability check: the average is updated only if the current signal variation remains within a specified delta (\texttt{baseline\_d}) of the previous baseline, ensuring that pulses do not skew the baseline estimation.

The \texttt{over\_check} entity receives this baseline and performs two critical functions:
\begin{itemize}
    \item \textbf{Thresholding:} It calculates a dynamic threshold (\texttt{baseline} - \texttt{thrd\_value\_rel}) and generates a bit-mask (\texttt{overthrd}) identifying which of the 8 samples in the current frame are active.
    \item \textbf{Conditioning:} It subtracts the baseline from the raw samples to produce \texttt{ADC\_check}, which is passed to the filtering stage.
\end{itemize}

\subsection{Real-Time Waveform Deconvolution (\texttt{filter})}
The \texttt{filter} wrapper encapsulates the core signal processing engines: \texttt{wiener\_filter} and \texttt{deconv\_filter}. These are implemented as parallel systolic \gls{fir} filters to handle the high data throughput: their implementation is handled using the Vivado LogiCORE$^{\mathrm{TM}}$ IP FIR Compiler~\cite{fir_compiler_guide}.

\begin{itemize}
    \item \textbf{Data Path:} The data flows sequentially from the Wiener filter (denoising) to the Deconvolution filter (ballistic deficit correction). Bit resizing logic is applied between stages to manage bit-growth and maintain the 16-bit sample width.
    \item \textbf{Configuration:} A finite state machine (FSM) manages the reloading of filter coefficients. The FSM transitions through \texttt{idle}, \texttt{load}, and \texttt{config} states, allowing dynamic reconfiguration of the \gls{fir} taps via an internal configuration FIFO (\texttt{fir\_config\_fifo}) without halting the rest of the system.
\end{itemize}

\subsection{Peak Finding(\texttt{peak})}
The \texttt{peak} entity identifies the maximum amplitude (charge) and its precise timing within the processed waveform.

\begin{itemize}
    \item \textbf{Pipeline:} A multi-stage pipeline (\texttt{peak\_find1} to \texttt{peak\_find3}) compares neighboring samples to isolate the local maximum.
    \item \textbf{Pattern Matching:} To filter out false positives, the \texttt{identify} process analyzes the vicinity of the maximum. It confirms a valid peak only if the samples strictly increase before and decrease after the maximum, effectively applying a shape constraint.
    \item \textbf{Output Generation:} When a peak is validated, the module outputs the \texttt{charge} and a fine-grained \texttt{hit\_time} (sample index within the frame).
\end{itemize}

\subsection{Output Logic}
Finally, the \texttt{TQ\_reco} top-level calculates the absolute timestamp \texttt{start\_time\_int} by combining the coarse system timestamp (\texttt{time\_cnt}), the pipeline delays, and the fine \texttt{hit\_time}. This information, together with charge and baseline values, is packed into a 128-bit \gls{tq} pair and written to the output FIFO.

\subsection{Resource Utilization}
Finally, the resource utilization of the implemented \gls{tq} reconstruction algorithm on the target Xilinx Kintex-7 XC7K325T \gls{fpga} is analyzed. The implementation results, summarized in \autoref{tab:usage}, highlight that the design is primarily constrained by the number of available DSP48E1 slices (840 total). 

% \begin{table}[htbp]
%     \centering
%     \begin{tabular}{l|ccccc}
%         \toprule
%         \textbf{Resource} & \textbf{Available} & \textbf{Tot. Util.} & \textbf{Tot. Util. (\%)} & \textbf{RTWD Util.} & \textbf{RTWD Util. (\%)} \\ 
%         \midrule
%         LUT & 203800 & 71494 & 35.08 & 6210 & 3.06 \\ 
%         LUTRAM & 64000 & 12937 & 20.21 & 3444 & 5.40 \\ 
%         FF & 407600 & 117219 & 28.76 & 26676 & 6.54 \\ 
%         BRAM & 445 & 389 & 87.42 & 12 & 2.70 \\ 
%         DSP & 840 & 624 & 74.28 & 624 & 74.28 \\ 
%         \bottomrule
%     \end{tabular}
%     \caption{Resource Utilization Summary}
%     \label{tab:usage}
% \end{table}

\begin{table}[htbp]
    \centering
    \begin{tblr}{
      column{even} = {c},
      column{3} = {c},
      column{5} = {c},
      hline{1,7} = {-}{0.08em},
      hline{2} = {-}{0.05em},
    }
    & \textbf{Available} & \textbf{Tot. Util.} & \textbf{Tot. Util. (\%)} & \textbf{RTWD Util.} & \textbf{RTWD Util. (\%)} \\
    \textbf{LUT}    & 203800             & 71494               & 35.08                    & 6210                & 3.06  \\
    \textbf{LUTRAM} & 64000              & 12937               & 20.21                    & 3444                & 5.40  \\
    \textbf{FF}     & 407600             & 117219              & 28.76                    & 26676               & 6.54  \\
    \textbf{BRAM}   & 445                & 389                 & 87.42                    & 12                  & 2.70  \\
    \textbf{DSP}    & 840                & 624                 & 74.28                    & 624                 & 74.28                    
    \end{tblr}
    
    \caption{Summary of resource utilization: \glsxtrfull{lut}, \glsxtrfull{lutram}, \glsxtrfull{ff}, \glsxtrfull{bram}, \glsxtrfull{dsp}, The table indicates that the main consumption of \glspl{dsp} stems from the RTWD implementation. The scarcity of DSP48E1 resources on the FPGA restricts parallel processing, which in turn limits the scope of possible FIR filter implementations.}
    \label{tab:usage}
\end{table}

The parallel implementation of the \gls{fir} filters is the dominant consumer of these resources: the 11-tap Wiener filter and the 7-tap Deconvolution filter, which process 8 samples per clock cycle to match the \gls{adc} throughput, require a significant density of multiply accumulation operations. However, the optimized RTL implementation ensures that \gls{dsp} usage remains within the device's limits.

\acknowledgments{
The authors acknowledge financial support by the Italian Ministry of University and Research (MUR), Call for tender No. 104/2022 (PRIN 2022) funded by the European Union (NextGenerationEU),\; Project "2022NAE9AJ - Getting ready to capture an exploding star", \\CUP C53D23001530006.
\\
\\The authors appreciate the valuable support and assistance provided by Hu Jun (IHEP) throughout the course of this work.
}

\bibliographystyle{JHEP}
\bibliography{biblio.bib}

@book{DFD,
    author = {T.W. Parks and C.S. Burros}, 
    title = {Digital Filter Design},
    year = {1987},
    publisher = {Wiley-Interscience,New York},
}

@article{Grassi_2018,
    doi = {10.1088/1748-0221/13/02/P02008},
    url = {https://dx.doi.org/10.1088/1748-0221/13/02/P02008},
    year = {2018},
    month = {2},
    publisher = {},
    volume = {13},
    number = {02},
    pages = {P02008},
    author = {M. Grassi and M. Montuschi and M. Baldoncini and F. Mantovani and B. Ricci and G. Andronico and V. Antonelli and M. Bellato and E. Bernieri and A. Brigatti and R. Brugnera and A. Budano and M. Buscemi and S. Bussino and R. Caruso and D. Chiesa and D. Corti and F. Dal Corso and X.F. Ding and S. Dusini and A. Fabbri and G. Fiorentini and R. Ford and A. Formozov and G. Galet and A. Garfagnini and M. Giammarchi and A. Giaz and A. Insolia and R. Isocrate and I. Lippi and F. Longhitano and D. Lo Presti and P. Lombardi and F. Marini and S.M. Mari and C. Martellini and E. Meroni and M. Mezzetto and L. Miramonti and S. Monforte and M. Nastasi and F. Ortica and A. Paoloni and S. Parmeggiano and D. Pedretti and N. Pelliccia and R. Pompilio and E. Previtali and G. Ranucci and A.C. Re and A. Romani and P. Saggese and G. Salamanna and F. H. Sawy and G. Settanta and M. Sisti and C. Sirignano and M. Spinetti and L. Stanco and V. Strati and G. Verde and L. Votano},
    title = {Charge reconstruction in large-area photomultipliers},
    journal = {Journal of Instrumentation}
}

@book{wiener,
    author = {S.W.Smith},
    title = {The Scientist and Engineer's Guide to Digital Signal Processing},
    publisher = {California Technical Pub},
    year = {1998},
    pages = {300-310}
}

@book{vaseghi2008advanced,
    author = {Saeed V. Vaseghi},
    title = {Advanced digital signal processing and noise reduction},
    year = {2008},
    publisher = {John Wiley \& Sons},
    pages = {182-183}
}

@phdthesis{fmarini,
    author = {Filippo Marini},
    title = {Development and Testing of the large PMTs Front-End Electronics for the JUNO Experiment},
    school = {University of Padova},
    year = {2021},
    url = {https://hdl.handle.net/11577/3443916}
}

@article{An_2016,
    doi = {10.1088/0954-3899/43/3/030401},
    url = {https://dx.doi.org/10.1088/0954-3899/43/3/030401},
    year = {2016},
    month = {02},
    publisher = {IOP Publishing},
    volume = {43},
    number = {3},
    pages = {030401},
    author = {Fengpeng An and Guangpeng An and Qi An and Vito Antonelli and Eric Baussan and John Beacom and Leonid Bezrukov and Simon Blyth and Riccardo Brugnera and Margherita Buizza Avanzini and Jose Busto and Anatael Cabrera and Hao Cai and Xiao Cai and Antonio Cammi and Guofu Cao and Jun Cao and Yun Chang and Shaomin Chen and Shenjian Chen and Yixue Chen and Davide Chiesa and Massimiliano Clemenza and Barbara Clerbaux and Janet Conrad and Davide D’Angelo and Hervé De Kerret and Zhi Deng and Ziyan Deng and Yayun Ding and Zelimir Djurcic and Damien Dornic and Marcos Dracos and Olivier Drapier and Stefano Dusini and Stephen Dye and Timo Enqvist and Donghua Fan and Jian Fang and Laurent Favart and Richard Ford and Marianne Göger-Neff and Haonan Gan and Alberto Garfagnini and Marco Giammarchi and Maxim Gonchar and Guanghua Gong and Hui Gong and Michel Gonin and Marco Grassi and Christian Grewing and Mengyun Guan and Vic Guarino and Gang Guo and Wanlei Guo and Xin-Heng Guo and Caren Hagner and Ran Han and Miao He and Yuekun Heng and Yee Hsiung and Jun Hu and Shouyang Hu and Tao Hu and Hanxiong Huang and Xingtao Huang and Lei Huo and Ara Ioannisian and Manfred Jeitler and Xiangdong Ji and Xiaoshan Jiang and Cécile Jollet and Li Kang and Michael Karagounis and Narine Kazarian and Zinovy Krumshteyn and Andre Kruth and Pasi Kuusiniemi and Tobias Lachenmaier and Rupert Leitner and Chao Li and Jiaxing Li and Weidong Li and Weiguo Li and Xiaomei Li and Xiaonan Li and Yi Li and Yufeng Li and Zhi-Bing Li and Hao Liang and Guey-Lin Lin and Tao Lin and Yen-Hsun Lin and Jiajie Ling and Ivano Lippi and Dawei Liu and Hongbang Liu and Hu Liu and Jianglai Liu and Jianli Liu and Jinchang Liu and Qian Liu and Shubin Liu and Shulin Liu and Paolo Lombardi and Yongbing Long and Haoqi Lu and Jiashu Lu and Jingbin Lu and Junguang Lu and Bayarto Lubsandorzhiev and Livia Ludhova and Shu Luo and  Vladimir Lyashuk and Randolph Möllenberg and Xubo Ma and Fabio Mantovani and Yajun Mao and Stefano M Mari and William F McDonough and Guang Meng and Anselmo Meregaglia and Emanuela Meroni and Mauro Mezzetto and Lino Miramonti and  Thomas Mueller and Dmitry Naumov and Lothar Oberauer and Juan Pedro Ochoa-Ricoux and Alexander Olshevskiy and Fausto Ortica and Alessandro Paoloni and Haiping Peng and  Jen-Chieh Peng and Ezio Previtali and Ming Qi and Sen Qian and Xin Qian and Yongzhong Qian and Zhonghua Qin and Georg Raffelt and Gioacchino Ranucci and Barbara Ricci and Markus Robens and Aldo Romani and Xiangdong Ruan and Xichao Ruan and Giuseppe Salamanna and Mike Shaevitz and  Valery Sinev and Chiara Sirignano and Monica Sisti and Oleg Smirnov and Michael Soiron and Achim Stahl and Luca Stanco and Jochen Steinmann and Xilei Sun and Yongjie Sun and Dmitriy Taichenachev and Jian Tang and Igor Tkachev and Wladyslaw Trzaska and Stefan van Waasen and Cristina Volpe and Vit Vorobel and Lucia Votano and Chung-Hsiang Wang and Guoli Wang and Hao Wang and Meng Wang and Ruiguang Wang and Siguang Wang and Wei Wang and Yi Wang and Yi Wang and Yifang Wang and Zhe Wang and Zheng Wang and Zhigang Wang and Zhimin Wang and Wei Wei and Liangjian Wen and Christopher Wiebusch and Björn Wonsak and Qun Wu and Claudia-Elisabeth Wulz and Michael Wurm and Yufei Xi and Dongmei Xia and Yuguang Xie and  Zhi-zhong Xing and Jilei Xu and Baojun Yan and Changgen Yang and Chaowen Yang and Guang Yang and Lei Yang and Yifan Yang and Yu Yao and Ugur Yegin and Frédéric Yermia and Zhengyun You and Boxiang Yu and Chunxu Yu and Zeyuan Yu and Sandra Zavatarelli and Liang Zhan and Chao Zhang and Hong-Hao Zhang and Jiawen Zhang and Jingbo Zhang and Qingmin Zhang and Yu-Mei Zhang and Zhenyu Zhang and Zhenghua Zhao and Yangheng Zheng and Weili Zhong and Guorong Zhou and Jing Zhou and Li Zhou and Rong Zhou and Shun Zhou and Wenxiong Zhou and Xiang Zhou and Yeling Zhou and Yufeng Zhou and Jiaheng Zou},
    title = {Neutrino physics with JUNO},
    journal = {Journal of Physics G: Nuclear and Particle Physics}
}

@misc{catiroc,
    title={CATIROC: an integrated chip for neutrino experiments using photomultiplier tubes}, 
    author={Selma Conforti and Mariangela Settimo and Cayetano Santos and Clément Bordereau and Anatael Cabrera and Stéphane Callier and Cédric Cerna and Christophe De La Taille and Frédéric Druillole and Frédéric Dulucq and Victor Lebrin and Frédéric Lefèvre and Gisèle Martin-Chassard and Frédéric Perrot and Abdel Rebii and Louis-Marie Rigalleau and Nathalie Seguin-Moreau},
    year={2021},
    eprint={2012.01565},
    archivePrefix={arXiv},
    primaryClass={physics.ins-det},
    url={https://arxiv.org/abs/2012.01565}
}

@article{ipbus,
doi = {10.1088/1748-0221/10/02/C02019},
url = {https://dx.doi.org/10.1088/1748-0221/10/02/C02019},
year = {2015},
month = {02},
publisher = {},
volume = {10},
number = {02},
pages = {C02019},
author = {C. Ghabrous Larrea and K. Harder and D. Newbold and D. Sankey and A. Rose and A. Thea and T. Williams},
title = {IPbus: a flexible Ethernet-based control system for xTCA hardware},
journal = {Journal of Instrumentation}
}

@ARTICLE{Abusleme2022,
	author = {Abusleme, Angel and Adam, Thomas and Ahmad, Shakeel and Ahmed, Rizwan and Aiello, Sebastiano and Akram, Muhammad and Aleem, Abid and Alexandros, Tsagkarakis and An, Fengpeng and An, Qi and Andronico, Giuseppe and Anfimov, Nikolay and Antonelli, Vito and Antoshkina, Tatiana and Asavapibhop, Burin and de André, João Pedro Athayde Marcondes and Auguste, Didier and Bai, Weidong and Balashov, Nikita and Baldini, Wander and Barresi, Andrea and Basilico, Davide and Baussan, Eric and Bellato, Marco and Bergnoli, Antonio and Birkenfeld, Thilo and Blin, Sylvie and Blum, David and Blyth, Simon and Bolshakova, Anastasia and Bongrand, Mathieu and Bordereau, Clément and Breton, Dominique and Brigatti, Augusto and Brugnera, Riccardo and Bruno, Riccardo and Budano, Antonio and Busto, Jose and Butorov, Ilya and Cabrera, Anatael and Caccianiga, Barbara and Cai, Hao and Cai, Xiao and Cai, Yanke and Cai, Zhiyan and Callegari, Riccardo and Cammi, Antonio and Campeny, Agustin and Cao, Chuanya and Cao, Guofu and Cao, Jun and Caruso, Rossella and Cerna, Cédric and Chan, Chi and Chang, Jinfan and Chang, Yun and Chen, Guoming and Chen, Pingping and Chen, Po-An and Chen, Shaomin and Chen, Xurong and Chen, Yixue and Chen, Yu and Chen, Zhiyuan and Chen, Zikang and Cheng, Jie and Cheng, Yaping and Cheng, Yu Chin and Chetverikov, Alexey and Chiesa, Davide and Chimenti, Pietro and Chukanov, Artem and Claverie, Gérard and Clementi, Catia and Clerbaux, Barbara and Colomer Molla, Marta and Conforti Di Lorenzo, Selma and Corti, Daniele and Corso, Flavio Dal and Dalager, Olivia and De La Taille, Christophe and Deng, Zhi and Deng, Ziyan and Depnering, Wilfried and Diaz, Marco and Ding, Xuefeng and Ding, Yayun and Dirgantara, Bayu and Dmitrievsky, Sergey and Dohnal, Tadeas and Dolzhikov, Dmitry and Donchenko, Georgy and Dong, Jianmeng and Doroshkevich, Evgeny and Dracos, Marcos and Druillole, Frédéric and Du, Ran and Du, Shuxian and Dusini, Stefano and Dvorak, Martin and Enqvist, Timo and Enzmann, Heike and Fabbri, Andrea and Fan, Donghua and Fan, Lei and Fang, Jian and Fang, Wenxing and Fargetta, Marco and Fedoseev, Dmitry and Fei, Zhengyong and Feng, Li-Cheng and Feng, Qichun and Ford, Richard and Fournier, Amélie and Gan, Haonan and Gao, Feng and Garfagnini, Alberto and Gavrikov, Arsenii and Giammarchi, Marco and Giudice, Nunzio and Gonchar, Maxim and Gong, Guanghua and Gong, Hui and Gornushkin, Yuri and Göttel, Alexandre and Grassi, Marco and Gromov, Vasily and Gu, Minghao and Gu, Xiaofei and Gu, Yu and Guan, Mengyun and Guan, Yuduo and Guardone, Nunzio and Guo, Cong and Guo, Jingyuan and Guo, Wanlei and Guo, Xinheng and Guo, Yuhang and Hackspacher, Paul and Hagner, Caren and Han, Ran and Han, Yang and He, Miao and He, Wei and Heinz, Tobias and Hellmuth, Patrick and Heng, Yuekun and Herrera, Rafael and Hor, YuenKeung and Hou, Shaojing and Hsiung, Yee and Hu, Bei-Zhen and Hu, Hang and Hu, Jianrun and Hu, Jun and Hu, Shouyang and Hu, Tao and Hu, Yuxiang and Hu, Zhuojun and Huang, Guihong and Huang, Hanxiong and Huang, Kaixuan and Huang, Wenhao and Huang, Xin and Huang, Xingtao and Huang, Yongbo and Hui, Jiaqi and Huo, Lei and Huo, Wenju and Huss, Cédric and Hussain, Safeer and Ioannisian, Ara and Isocrate, Roberto and Jelmini, Beatrice and Jeria, Ignacio and Ji, Xiaolu and Jia, Huihui and Jia, Junji and Jian, Siyu and Jiang, Di and Jiang, Wei and Jiang, Xiaoshan and Jing, Xiaoping and Jollet, Cécile and Joutsenvaara, Jari and Kalousis, Leonidas and Kampmann, Philipp and Kang, Li and Karaparambil, Rebin and Kazarian, Narine and Khatun, Amina and Khosonthongkee, Khanchai and Korablev, Denis and Kouzakov, Konstantin and Krasnoperov, Alexey and Kutovskiy, Nikolay and Kuusiniemi, Pasi and Lachenmaier, Tobias and Landini, Cecilia and Leblanc, Sébastien and Lebrin, Victor and Lefevre, Frederic and Lei, Ruiting and Leitner, Rupert and Leung, Jason and Li, Daozheng and Li, Demin and Li, Fei and Li, Fule and Li, Gaosong and Li, Huiling and Li, Mengzhao and Li, Min and Li, Nan and Li, Nan and Li, Qingjiang and Li, Ruhui and Li, Rui and Li, Shanfeng and Li, Tao and Li, Teng and Li, Weidong and Li, Weiguo and Li, Xiaomei and Li, Xiaonan and Li, Xinglong and Li, Yi and Li, Yichen and Li, Yufeng and Li, Zepeng and Li, Zhaohan and Li, Zhibing and Li, Ziyuan and Li, Zonghai and Liang, Hao and Liang, Hao and Liao, Jiajun and Limphirat, Ayut and Lin, Guey-Lin and Lin, Shengxin and Lin, Tao and Ling, Jiajie and Lippi, Ivano and Liu, Fang and Liu, Haidong and Liu, Haotian and Liu, Hongbang and Liu, Hongjuan and Liu, Hongtao and Liu, Hui and Liu, Jianglai and Liu, Jinchang and Liu, Min and Liu, Qian and Liu, Qin and Liu, Runxuan and Liu, Shubin and Liu, Shulin and Liu, Xiaowei and Liu, Xiwen and Liu, Yan and Liu, Yunzhe and Lokhov, Alexey and Lombardi, Paolo and Lombardo, Claudio and Loo, Kai and Lu, Chuan and Lu, Haoqi and Lu, Jingbin and Lu, Junguang and Lu, Shuxiang and Lubsandorzhiev, Bayarto and Lubsandorzhiev, Sultim and Ludhova, Livia and Lukanov, Arslan and Luo, Daibin and Luo, Fengjiao and Luo, Guang and Luo, Shu and Luo, Wuming and Luo, Xiaojie and Lyashuk, Vladimir and Ma, Bangzheng and Ma, Bing and Ma, Qiumei and Ma, Si and Ma, Xiaoyan and Ma, Xubo and Maalmi, Jihane and Mai, Jingyu and Malyshkin, Yury and Mandujano, Roberto Carlos and Mantovani, Fabio and Manzali, Francesco and Mao, Xin and Mao, Yajun and Mari, Stefano M. and Marini, Filippo and Martellini, Cristina and Martin-Chassard, Gisele and Martini, Agnese and Mayer, Matthias and Mayilyan, Davit and Mednieks, Ints and Meng, Yue and Meregaglia, Anselmo and Meroni, Emanuela and Meyhöfer, David and Mezzetto, Mauro and Miller, Jonathan and Miramonti, Lino and Montini, Paolo and Montuschi, Michele and Müller, Axel and Nastasi, Massimiliano and Naumov, Dmitry V. and Naumova, Elena and Navas-Nicolas, Diana and Nemchenok, Igor and Nguyen Thi, Minh Thuan and Ning, Feipeng and Ning, Zhe and Nunokawa, Hiroshi and Oberauer, Lothar and Ochoa-Ricoux, Juan Pedro and Olshevskiy, Alexander and Orestano, Domizia and Ortica, Fausto and Othegraven, Rainer and Paoloni, Alessandro and Parmeggiano, Sergio and Pei, Yatian and Pelicci, Luca and Pelliccia, Nicomede and Peng, Anguo and Peng, Haiping and Peng, Yu and Peng, Zhaoyuan and Perrot, Frédéric and Petitjean, Pierre-Alexandre and Petrucci, Fabrizio and Pilarczyk, Oliver and Piñeres Rico, Luis Felipe and Popov, Artyom and Poussot, Pascal and Previtali, Ezio and Qi, Fazhi and Qi, Ming and Qian, Sen and Qian, Xiaohui and Qian, Zhen and Qiao, Hao and Qin, Zhonghua and Qiu, Shoukang and Ranucci, Gioacchino and Raper, Neill and Re, Alessandra and Rebber, Henning and Rebii, Abdel and Redchuk, Mariia and Ren, Bin and Ren, Jie and Ricci, Barbara and Rifai, Mariam and Roche, Mathieu and Rodphai, Narongkiat and Romani, Aldo and Roskovec, Bedřich and Ruan, Xichao and Rybnikov, Arseniy and Sadovsky, Andrey and Saggese, Paolo and Sanfilippo, Simone and Sangka, Anut and Sawangwit, Utane and Sawatzki, Julia and Schever, Michaela and Schwab, Cédric and Schweizer, Konstantin and Selyunin, Alexandr and Serafini, Andrea and Settanta, Giulio and Settimo, Mariangela and Shao, Zhuang and Sharov, Vladislav and Shaydurova, Arina and Shi, Jingyan and Shi, Yanan and Shutov, Vitaly and Sidorenkov, Andrey and Šimkovic, Fedor and Sirignano, Chiara and Siripak, Jaruchit and Sisti, Monica and Slupecki, Maciej and Smirnov, Mikhail and Smirnov, Oleg and Sogo-Bezerra, Thiago and Sokolov, Sergey and Songwadhana, Julanan and Soonthornthum, Boonrucksar and Sotnikov, Albert and Šrámek, Ondřej and Sreethawong, Warintorn and Stahl, Achim and Stanco, Luca and Stankevich, Konstantin and Štefánik, Dušan and Steiger, Hans and Steinmann, Jochen and Sterr, Tobias and Stock, Matthias Raphael and Strati, Virginia and Studenikin, Alexander and Su, Jun and Sun, Shifeng and Sun, Xilei and Sun, Yongjie and Sun, Yongzhao and Sun, Zhengyang and Suwonjandee, Narumon and Szelezniak, Michal and Tang, Jian and Tang, Qiang and Tang, Quan and Tang, Xiao and Theisen, Eric and Tietzsch, Alexander and Tkachev, Igor and Tmej, Tomas and Torri, Marco Danilo Claudio and Treskov, Konstantin and Triossi, Andrea and Troni, Giancarlo and Trzaska, Wladyslaw and Tuve, Cristina and Ushakov, Nikita and Vedin, Vadim and Verde, Giuseppe and Vialkov, Maxim and Viaud, Benoit and Vollbrecht, Cornelius Moritz and Volpe, Cristina and von Sturm, Katharina and Vorobel, Vit and Voronin, Dmitriy and Votano, Lucia and Walker, Pablo and Wang, Caishen and Wang, Chung-Hsiang and Wang, En and Wang, Guoli and Wang, Jian and Wang, Jun and Wang, Lu and Wang, Meifen and Wang, Meng and Wang, Ruiguang and Wang, Siguang and Wang, Wei and Wang, Wenshuai and Wang, Xi and Wang, Xiangyue and Wang, Yangfu and Wang, Yaoguang and Wang, Yi and Wang, Yi and Wang, Yifang and Wang, Yuanqing and Wang, Yuman and Wang, Zhe and Wang, Zheng and Wang, Zhimin and Wang, Zongyi and Watcharangkool, Apimook and Wei, Wei and Wei, Wei and Wei, Wenlu and Wei, Yadong and Wen, Kaile and Wen, Liangjian and Wiebusch, Christopher and Wong, Steven Chan-Fai and Wonsak, Bjoern and Wu, Diru and Wu, Qun and Wu, Zhi and Wurm, Michael and Wurtz, Jacques and Wysotzki, Christian and Xi, Yufei and Xia, Dongmei and Xiao, Xiang and Xie, Xiaochuan and Xie, Yuguang and Xie, Zhangquan and Xin, Zhao and Xing, Zhizhong and Xu, Benda and Xu, Cheng and Xu, Donglian and Xu, Fanrong and Xu, Hangkun and Xu, Jilei and Xu, Jing and Xu, Meihang and Xu, Yin and Xu, Yu and Yan, Baojun and Yan, Taylor and Yan, Wenqi and Yan, Xiongbo and Yan, Yupeng and Yang, Changgen and Yang, Chengfeng and Yang, Huan and Yang, Jie and Yang, Lei and Yang, Xiaoyu and Yang, Yifan and Yao, Haifeng and Ye, Jiaxuan and Ye, Mei and Ye, Ziping and Yermia, Frédéric and Yin, Na and You, Zhengyun and Yu, Boxiang and Yu, Chiye and Yu, Chunxu and Yu, Hongzhao and Yu, Miao and Yu, Xianghui and Yu, Zeyuan and Yu, Zezhong and Yuan, Cenxi and Yuan, Chengzhuo and Yuan, Ying and Yuan, Zhenxiong and Yue, Baobiao and Zafar, Noman and Zavadskyi, Vitalii and Zeng, Shan and Zeng, Tingxuan and Zeng, Yuda and Zhan, Liang and Zhang, Aiqiang and Zhang, Bin and Zhang, Binting and Zhang, Feiyang and Zhang, Guoqing and Zhang, Honghao and Zhang, Jialiang and Zhang, Jiawen and Zhang, Jie and Zhang, Jin and Zhang, Jingbo and Zhang, Jinnan and Zhang, Mohan and Zhang, Peng and Zhang, Qingmin and Zhang, Shiqi and Zhang, Shu and Zhang, Tao and Zhang, Xiaomei and Zhang, Xin and Zhang, Xuantong and Zhang, Xueyao and Zhang, Yinhong and Zhang, Yiyu and Zhang, Yongpeng and Zhang, Yu and Zhang, Yuanyuan and Zhang, Yumei and Zhang, Zhenyu and Zhang, Zhijian and Zhao, Fengyi and Zhao, Jie and Zhao, Rong and Zhao, Runze and Zhao, Shujun and Zheng, Dongqin and Zheng, Hua and Zheng, Yangheng and Zhong, Weirong and Zhou, Jing and Zhou, Li and Zhou, Nan and Zhou, Shun and Zhou, Tong and Zhou, Xiang and Zhu, Jiang and Zhu, Jingsen and Zhu, Kangfu and Zhu, Kejun and Zhu, Zhihang and Zhuang, Bo and Zhuang, Honglin and Zong, Liang and Zou, Jiaheng},
	title = {Mass testing and characterization of 20-inch PMTs for JUNO},
	year = {2022},
	journal = {European Physical Journal C},
	volume = {82},
	number = {12},
	doi = {10.1140/epjc/s10052-022-11002-8},
	url = {https://www.scopus.com/inward/record.uri?eid=2-s2.0-85138851306&doi=10.1140%2fepjc%2fs10052-022-11002-8&partnerID=40&md5=3bcfd3e19dc913564d9d7d6fbc6bc27d},
	type = {Article},
	publication_stage = {Final},
	source = {Scopus},
	note = {Cited by: 15; All Open Access, Gold Open Access}
}

@article{remez,
    author = {Fraser, W.},
    title = {A Survey of Methods of Computing Minimax and Near-Minimax Polynomial Approximations for Functions of a Single Independent Variable},
    year = {1965},
    issue_date = {July 1965},
    publisher = {Association for Computing Machinery},
    address = {New York, NY, USA},
    volume = {12},
    number = {3},
    issn = {0004-5411},
    url = {https://doi.org/10.1145/321281.321282},
    doi = {10.1145/321281.321282},
    journal = {J. ACM},
    month = jul,
    pages = {295–314},
    numpages = {20}
}

@article{LAVITOLA2023168461,
title = {multi-PMT electronics system for Hyper-Kamiokande},
journal = {Nuclear Instruments and Methods in Physics Research Section A: Accelerators, Spectrometers, Detectors and Associated Equipment},
volume = {1054},
pages = {168461},
year = {2023},
issn = {0168-9002},
doi = {https://doi.org/10.1016/j.nima.2023.168461},
url = {https://www.sciencedirect.com/science/article/pii/S0168900223004515},
author = {Luigi Lavitola}
}

@article{ipbus-juno,
    author = "Triozzi, Riccardo and others",
    title = "{Implementation and performances of the IPbus protocol for the JUNO Large-PMT readout electronics}",
    eprint = "2302.10133",
    archivePrefix = "arXiv",
    primaryClass = "physics.ins-det",
    doi = "10.1016/j.nima.2023.168339",
    journal = "Nucl. Instrum. Meth. A",
    volume = "1053",
    pages = "168339",
    year = "2023"
}

@article{Liu_2023,
    doi = {10.1088/1748-0221/18/02/P02003},
    url = {https://doi.org/10.1088/1748-0221/18/02/P02003},
    year = {2023},
    month = {2},
    publisher = {IOP Publishing},
    volume = {18},
    number = {02},
    pages = {P02003},
    author = {Liu, Caimei and Li, Min and Wang, Zhimin and Hu, Jun and Anfimov, Nikolay and Fan, Lei and Garfagnini, Alberto and Gong, Guanghua and Hou, Shaojing and Ji, Xiaolu and Jiang, Xiaoshan and Korablev, Denis and Lachenmaier, Tobias and Ma, Si and Ma, Xiaoyan and Ning, Zhe and Olshevskiy, Alexander G. and Peng, Zhaoyuan and Qin, Zhonghua and Sterr, Tobias and Sun, Yunhua and Tietzsch, Alexander Felix and Wang, Jun and Wang, Wei and Wang, Yangfu and Wen, Kaile and Wonsak, Bjoern Soenke and Xie, Wan and Xu, Meihang and Yan, Xiongbo and Yang, Yifan and Zhao, Rong and Zhou, Tong and Zhu, Kejun},
    title = {Check on the features of potted 20-inch PMTs with 1F3 electronics prototype at Pan-Asia},
    journal = {Journal of Instrumentation},
}

@misc{liu2025juno20inchpmtelectronics,
      title={JUNO 20-inch PMT and electronics system characterization using large pulses of PMT dark counts at the Pan-Asia testing platform}, 
      author={Caimei Liu and Min Li and Narongkiat Rodphai and Zhimin Wang and Jun Hu and Nikolay Anfimov and Lei Fan and Alberto Garfagnini and Guanghua Gong and Shaojing Hou and Xiaolu Ji and Xiaoshan Jiang and Denis Korablev and Tobias Lachenmaier and Si Ma and Xiaoyan Ma and Zhe Ning and Alexander G. Olshevskiy and Zhaoyuan Peng and Zhonghua Qin and Tobias Sterr and Yunhua Sun and Alexander Felix Tietzsch and Jun Wang and Wei Wang and Yangfu Wang and Kaile Wen and Bjoern Soenke Wonsak and Wan Xie and Meihang Xu and Xiongbo Yan and Yifan Yang and Rong Zhao and Tong Zhou and Kejun Zhu and Jianmeng Dong and Pierre-Alexandre Petitjean and Barbara Clerbaux},
      year={2025},
      eprint={2506.21179},
      archivePrefix={arXiv},
      primaryClass={physics.ins-det},
      url={https://arxiv.org/abs/2506.21179}, 
}

@manual{caen,
    key = {DT5810},
    organization = {CAEN},
    title = {Fast Digital Detector Emulator DT5810},
    doi = {\href{https://www.caen.it/products/dt5810}{DT5810}},
    urldate = {2025-06-11}
}

@manual{scipy,
  author = {The SciPy community},
  title = {Signal processing (scipy.signal)},
  doi = {\href{https://docs.scipy.org/doc/scipy/reference/signal.html}{Signal processing (scipy.signal)}},
  year = {2008},
  urldate = {2025-04-12}
}

@article{McSweeney01042006,
    author = {Laura A. McSweeney},
    title = {Comparison of periodogram tests},
    journal = {Journal of Statistical Computation and Simulation},
    volume = {76},
    number = {4},
    pages = {357--369},
    year = {2006},
    publisher = {Taylor \& Francis},
    doi = {10.1080/10629360500107618},
    URL = {https://doi.org/10.1080/10629360500107618},
    eprint = {https://doi.org/10.1080/10629360500107618}
}

@article{IceCube:2016zyt,
    author = "Aartsen, M. G. and others",
    collaboration = "IceCube",
    title = "{The IceCube Neutrino Observatory: Instrumentation and Online Systems}",
    eprint = "1612.05093",
    archivePrefix = "arXiv",
    primaryClass = "astro-ph.IM",
    doi = "10.1088/1748-0221/12/03/P03012",
    journal = "JINST",
    volume = "12",
    number = "03",
    pages = "P03012",
    year = "2017",
    note = "[Erratum: JINST 19, E05001 (2024)]"
}

@article{KM3NeT:2022pnv,
    author = "Aiello, S. and others",
    collaboration = "KM3NeT",
    title = "{The KM3NeT multi-PMT optical module}",
    eprint = "2203.10048",
    archivePrefix = "arXiv",
    primaryClass = "astro-ph.IM",
    doi = "10.1088/1748-0221/17/07/P07038",
    journal = "JINST",
    volume = "17",
    pages = "P07038",
    year = "2022"
}

@manual{fir_compiler_guide,
    title = {FIR Compiler LogiCORE IP Product Guide ( PG149)},
    doi = {\href{https://docs.amd.com/r/en-US/pg149-fir-compiler/Introduction}{FIR Compiler}},
    author = {AMD}
}

@article{Zhang_2019,
   title={Comparison on PMT waveform reconstructions with JUNO prototype},
   volume={14},
   ISSN={1748-0221},
   url={http://dx.doi.org/10.1088/1748-0221/14/08/T08002},
   DOI={10.1088/1748-0221/14/08/t08002},
   number={08},
   journal={Journal of Instrumentation},
   publisher={IOP Publishing},
   author={Zhang, H.Q. and Wang, Z.M. and Zhang, Y.P. and Huang, Y.B. and Luo, F.J. and Zhang, P. and Zhang, C.C. and Xu, M.H. and Liu, J.C. and Heng, Y.K. and Yang, C.G. and Jiang, X.S. and Li, F. and Ye, M. and Chen, H.S.},
   year={2019},
   month=aug, pages={T08002–T08002} }
\end{document}